\documentclass[final,1p]{elsarticle}
\pdfoutput=1 
\usepackage[T1]{fontenc}
\usepackage[british]{babel}
\usepackage[utf8]{inputenc}
\usepackage[sf,SF]{subfigure}
\usepackage{lmodern}
\usepackage{makeidx}
\usepackage{multicol}
\usepackage{footmisc}
\usepackage{amsmath}
\usepackage{amssymb}
\usepackage{color}
\usepackage{url}
\usepackage{graphicx}
\usepackage{booktabs}   
\usepackage{centernot}
\usepackage{cleveref}

\crefname{chapter}{Chapter}{Chapters}
\crefname{section}{Section}{Sections}
\crefname{appendix}{Appendix}{Appendices}
\crefname{subsection}{Section}{Sections}
\crefname{subsubsection}{Section}{Sections}
\crefname{equation}{Equation}{Equations}
\crefname{figure}{Figure}{Figures}
\crefname{table}{Table}{Tables}
\crefname{subfigure}{Figure}{Figures}
\crefname{listing}{Listing}{Listings}

\newcommand{\x}{$\times$}
\newcommand{\eg}{e.g.\ }
\newcommand{\ie}{i.e.\ }
\newcommand{\etal}{et al.\ }

\renewcommand{\v}[1]{\ensuremath{\mathbf{#1}}} 
\newcommand{\gv}[1]{\ensuremath{\mbox{\boldmath$ #1 $}}}  
\newcommand{\pd}[2]{\frac{\partial #1}{\partial #2}} 
\newcommand{\grad}[1]{\gv{\nabla} #1} 
\renewcommand{\div}[1]{\gv{\nabla} \cdot #1} 

\journal{Journal of Computational Physics}

\begin{document}

\begin{frontmatter}

\title{A robust method for calculating interface curvature and normal vectors
using an extracted local level set}
\author[sintef,ntnu]{Åsmund Ervik\corref{cor1}}
\ead{asmund.ervik@sintef.no}
\cortext[cor1]{Corresponding author}
\author[ntnu]{Karl Yngve Lervåg}
\ead{karl.y.lervag@ntnu.no}
\author[sintef]{Svend Tollak Munkejord}
\ead{svend.t.munkejord@sintef.no}
\address[sintef]{SINTEF Energy Research,
P.O. Box 4761 Sluppen, NO-7465 Trondheim, Norway}
\address[ntnu]{NTNU, Department of Energy and Process Engineering,
Kolbjørn Hejes v 1B, NO-7491 Trondheim, Norway}

\begin{abstract}
  The level-set method is a popular interface tracking method in two-phase
  flow simulations. An often-cited reason for using it is that the method
  naturally handles topological changes in the interface, e.g. merging drops,
  due to the implicit formulation. It
  is also said that the interface curvature and normal vectors are easily
  calculated. This last point is not, however, the case in the moments during
  a topological change, as several authors have already pointed out.
  Various methods have been employed to circumvent the problem. In this paper,
  we present a new such method which retains the implicit
  level-set representation of the surface and handles general interface
  configurations. It is demonstrated that the method extends easily to 3D.
  The method is validated on static interface configurations,
  and then applied to two-phase flow simulations where the method outperforms
  the standard method and the results agree well with experiments.

\end{abstract}



\end{frontmatter}


\section{Introduction}
Investigations of droplet collision phenomena have a long tradition in the study
of multiphase flow, dating back to Lord Rayleigh \cite{rayleigh} who in 1879
noted that a raindrop can bounce off a pool, and to Worthington
\cite{worthington} who in 1876 studied among other things the central jet that
now bears his name. The early work predates the rise of computational studies,
and consists of experimental studies that enabled a separation of the flow
patterns into various regimes characterized by \eg the Weber number and
Ohnesorge number.  A case which has long been the focus of study is that of
a single droplet of one liquid, immersed in some other gas or liquid, and which
collides with a deep pool of the first liquid. This could be \eg a raindrop
falling onto a pond, or a droplet of Liquefied Natural Gas (LNG) merging with
a pool of LNG in a liquefaction heat exchanger, so the case is interesting also
from an industry point of view.  Such a system may seem simple at first, but
experimental and numerical studies have shown that varied phenomena such as
coalescence, bouncing, jetting and partial merging occur. The system is also not
fully understood yet; as an example, Thoroddsen \etal \cite{droplet-turbulence}
have recently shown  that for high impact velocities a turbulent boundary layer
forms between the droplet and the pool after they merge.

In order to study such a case using computer simulations, it is necessary
to use a precise interface-tracking method to capture the physics before,
during and after the collision. The Level-Set Method (LSM) is a popular
choice for interface tracking in studies of collisions, since its implicit
formulation means that the method can handle the topological change which
occurs when two interfaces merge. The LSM is very general, and apart from
fluid dynamics it has been used for modelling such diverse phenomena as
tumor growth \cite{macklin}, wildland fire propagation \cite{wildfire} and
computer RAM production \cite{ram}.  For a good introduction to the LSM,
see \eg \cite{osher-fedkiw}. The LSM originated from the seminal article by
Osher and Sethian \cite{founding}.

In two-phase flow simulations using the LSM, accurate interface curvature and
normal vector information is vital in order to get good results. Standard
methods exist for calculating these geometric quantities, but they fail when the
interface topology changes, \eg when two drops collide and merge.  Several
approaches have been used to remedy this flaw.  The first approach to this
problem is described by Smereka in \cite{smereka}. He describes the problem
briefly, and increases the numerical smoothing in the curvature discretization
to lessen the effect. This is not an optimal solution, and Smereka notes on one
of the simulations with merging interfaces that ``most of the area loss occurs
at the topology change''. Several non-smearing approaches have subsequently been
developed, by Macklin and Lowengrub \cite{macklin,macklin06}, by Salac and Lu
\cite{salac} and by Lervåg \cite{lervag-curv-conf,lervag-curv}.  The methods by
Macklin and Lowengrub and by Lervåg use curve fitting to obtain an accurate
representation of the interface, while the method by Salac and Lu extracts
several level-set functions each representing only a single body, and uses these
to calculate the curvature.

The present work proposes a new method, which is an extension of previous
methods, for calculating the curvature and normal vectors. The proposed
method is based on the method by Salac and Lu, but it handles more general
interface configurations and topological changes, as it considers only the
local area around a point.  The quality function introduced by Macklin and
Lowengrub is used to restrict the use of the proposed method to those areas
where it is needed, thus reducing the computational cost. As the proposed
method uses no curve fitting, it extends easily to three dimensions, as
demonstrated here. The proposed method is compared to the standard method
for demanding cases where the analytical curvature is known; for such a
case the proposed method gives errors of 1--2\% where the standard method
gives errors of $\mathcal{O}(1/\Delta x) > 100\%$. The proposed method is based
on the work of Ervik \cite{ervik}.

The outline of this work is as follows: In \cref{sec:theory}, the theory of
two-phase incompressible flow, the LSM and numerical methods are briefly reviewed. In
\cref{sec:lolex}, the proposed method is presented in detail. In
\cref{sec:geometric}, the method is validated on geometric test cases, and the
results are compared to other methods. In \cref{sec:simulations}, the results of
two-phase flow simulations using the current method are reported and compared to
experimental results. Finally, in \cref{sec:conclude}, some
concluding remarks are offered.


\section{The Level-Set Method and two-phase flow}
\label{sec:theory}
The LSM is one of the more successful interface-capturing methods used in
computational physics. Since its introduction by Osher and Sethian in
\cite{founding}, it has been used for numerous physical applications, as well as
in computer graphics. Perhaps the main virtue of the LSM is how intuitive it is;
in 2D it can easily be explained to anyone with a basic knowledge of
multivariate calculus. This simplicity stems from the implicitness of the LSM,
making the numerical implementation of the LSM relatively easy. The implicit
formulation also means that changes in the interface topology are handled
naturally. When comparing the LSM to other interface-tracking methods, such as
the front-tracking method \cite{fronttracking} where the interface is
represented by piecewise continuous functions, the simplicity becomes especially
clear.

The main disadvantage of the LSM, on the other hand, is that it is not
a conservative method. During the course of a simulation, a fraction of fluid
1 may be converted to fluid 2 in an unphysical fashion. Various methods have
been invented to circumvent this, \eg the HCR-2 reinitialization method
\cite{hartmann10}, so it is only a small effect presently. Interface-tracking
methods may be conservative; an example of this is the Volume-of-Fluid (VOF)
Method, but then they typically have other disadvantages. In the VOF method, for
instance, the advection equation cannot easily be solved, necessitating the use
of interface-reconstruction methods \cite{vof}. Recent efforts have attempted to
join the LSM and VOF in order to get the benefits of both methods; this approach
seems to be fairly successful \cite{hybridlsmvof}. In a similar spirit, recent
hybrid level-set/front-tracking methods have been developed \cite{shin11} that
retain the implicit definition of the interface while utilizing the
front-tracking method to improve mass conservation and to compute the
surface-tension forces in an accurate and robust manner. 

We give here the formal definition of the level-set function used in the LSM.
Let $\Gamma$ be the interface between
two fluids, \eg air and water, and $S$ be the computational domain where the
fluids are confined. To represent this interface, we define a
\emph{level-set function}
$\phi : S \to \mathbb{R}$ with the property
\begin{equation}
  \Gamma = \{\v{x} \; | \; \phi(\v{x}) = 0\} \rm{.}
  \label{eq:phi-defining}
\end{equation}
This only
defines the value of $\phi$ at the interface $\Gamma$, and not elsewhere. The
common choice here is a signed distance function. Thus $\phi$ is fully specified by
\begin{equation}
  \phi(\v{x}) =
\begin{cases}
  -\text{dist}(\v{x},\Gamma) & \text{ if } \v{x} \text{ is inside } \Gamma \rm{,}\\
  \hspace{8pt}\text{dist}(\v{x},\Gamma)  & \text{ if } \v{x} \text{ is outside } \Gamma \rm{.}
\end{cases}
\label{eq:phi}
\end{equation}
Here, the function $\text{dist}(\v{x},\Gamma)$ is the shortest distance from the
point $\v{x} \in S$ to the interface $\Gamma$.
With this definition of the level-set function, the normal vector to the
interface is given by
\begin{equation}
  \v{n} = \frac{\grad \phi}{|\grad\phi|} \rm{.}
  \label{eq:normal}
\end{equation}
From this, the curvature is calculated by the well-known formula
\begin{equation}
        \kappa = \div \v{n} = \div \left( \frac{\grad \phi}{|\grad \phi|} \right)\rm{.}
        \label{eq:curvature}
\end{equation}
With suitable discretizations of the derivatives involved, these quantities
are easy to calculate numerically. This is often quoted as one of the nice
features of the LSM, along with \eg the very natural way the method handles
topological changes \cite{locallsm}. In 2D, the standard discretization of
the curvature is (see \eg \cite{kang})
\begin{equation}
  \kappa = \frac{\phi_{xx} + \phi_{yy}}{(\phi_x^2+\phi_y^2+
    \epsilon)^{1/2}} - \frac{\phi_x^2\phi_{xx} + \phi_y^2\phi_{yy}
    + 2\phi_x\phi_y\phi_{xy}}{(\phi_x^2 + \phi_y^2+\epsilon)^{3/2}}
\label{eq:curv-disc}
\end{equation}
Here, \eg $\phi_{x}$ denotes the first derivative of $\phi$ in $x$-direction,
calculated using standard central differences. However, when curvature
and normal vector calculations are done during a change in the
interface topology, this approach fails; the error in curvature is of the order
$\mathcal{O}(1/\Delta x)$ \cite{macklin}. In \cite{smereka}, Smereka notes that
``One of the major advantages of level-set methods is their ability to easily
handle topological changes. However for this problem we have found this not
to be the case.'' It is this that the present method attempts to solve.

From the defining \cref{eq:phi},
$\phi$ is initialized at the start of
a simulation. For a given velocity field $\v{u}$, $\phi$ should be transported so
that the interface follows the flow. This is done by solving the advection
equation,
\begin{equation}
  \pd{\phi}{t} = v |\grad\phi| = -\v{u}\cdot\grad\phi \rm{.}
  \label{eq:advection}
\end{equation}
Here $v$ is the velocity normal to the interface, and $\v{u}$ is an
\emph{extrapolated} velocity field constructed using the method in
\cite{hansen}. This equation is not justified here, see \eg \cite{sussman}.

Solving this equation will result in transportation of the interface, but it
will also degrade the accuracy of the interface representation, as $\phi$ is
deformed from a signed distance function. To avoid this,
the level-set function is periodically \emph{reinitialized}. We follow here the
PDE-based approach introduced by Sussman,
Smereka and Osher \cite{sussman}, which consists in solving
\begin{equation}
  \pd{\phi}{\tau} + \text{sgn}(\phi)(|\grad\phi| - 1) = 0 \rm{.}
  \label{eq:reinit}
\end{equation}
Here $\tau$ is a pseudo-time which is not related to the physical
time in simulations. This approach is both computationally fast and accurate
when used as here with a narrow-band approach. The extrapolation of the velocity
field as used in \cref{eq:advection} above is performed by solving a similar
type of equation. These equations are solved using pseudo-CFL numbers
of 1.0 for the velocity extrapolation and 0.5 for the reinitialization. 
It is noted that a numerical solution of the reinitialization equation needs 
accurate normal vectors at the interface.

A useful property of these equations is that the characteristics originate at the
interface, meaning that solving the equations numerically for $N$ pseudo-time steps using
a CFL-number of $C$ will yield a correct signed distance function $C\cdot N$
space steps away from the interface. This has led to the use of narrow-band
methods, where the level-set function and other properties such as the curvature
are only calculated and used in a narrow band around the interface. This reduces
the computational time significantly.

In two-phase flow simulations, the LSM is coupled with the Navier-Stokes
equations,
\begin{align}
  \div\v{u} &= 0 \label{eq:incompressible} \rm{,} \\
        \pd{\v{u}}{t}+(\v{u}\cdot\grad)\v{u} &= - \frac{\grad{p}}{\rho} + \nu\grad^2\v{u} +
  \v{f} \rm{.}
        \label{eq:navier-stokes}
\end{align}
Here $\nu=\mu/\rho$ is the kinematic viscosity, while $\mu$ is the dynamic
viscosity, $\rho$ is the density, $\v{u}$ is the velocity field and $p$ is the
pressure. $\v{f}$ is any external force, such as gravity, and may be zero.

These equations hold for single-phase fluid flow, but can be extended to
two-phase flow using different methods. In the present work, the Ghost Fluid
Method (GFM) \cite{GFMNavier} is used. This method prescribes jump conditions
for \eg the pressure across the interface based on the interface properties. The
jump conditions used here are
\begin{align}
        \label{eq:jump1}
        [\v{u}] &= 0\rm{,} \\
        \label{eq:jump2}
  \left[ p \right] &= 2[\mu]\v{n}\cdot\grad\v{u}\cdot\v{n}+\sigma\kappa\rm{,} \\
        \left[\mu\grad\v{u}\right] &= [\mu]\bigg(
        (\v{n}\cdot\grad\v{u}\cdot\v{n})\v{n}\v{n} +
  (\v{n}\cdot\grad\v{u}\cdot\v{t})\v{n}\v{t}  \\[-0.7em] & \qquad \quad \; \hspace{2pt}
        - (\v{n}\cdot\grad\v{u}\cdot\v{t})\v{t}\v{n} \hspace{2pt} + \hspace{2pt}
        (\v{t}\cdot\grad\v{u}\cdot\v{t})\v{t}\v{t}\hspace{2pt}
        \bigg) \rm{,} \\
        \left[\grad p\right] &= 0\rm{.}
  \label{eq:jump4}
\end{align}
based on \cite{kang}. Here, $\v{t}$ is the tangent vector along the interface
and $[\cdot]$ denotes the jump across an interface, that is $[\mu] \equiv \mu^+
- \mu^-$.  Note that $\grad \v{u}$ and (e.g.) $\v{n}\v{t}$ are rank-2 tensors.
The pressure must also be decoupled from the velocity field in order to enable
a numerical solution of the Navier-Stokes equations; we use here the projection
method due to Chorin \cite{chorin}. This gives a Poisson equation for the
pressure which can be solved using freely available numerical libraries. The
PETSc library is used here \cite{petsc}.

In the present numerical implementation, SSP-RK schemes
\cite{kraaijevanger,ketcheson} are used for the time integration, while the
WENO method \cite{jiang} is used for the spatial discretizations. To
determine the time step dynamically, we use the CFL criterion given by Kang
\etal \cite{kang}.


\section{The Local Level-Set Extraction (LOLEX) Method}
\label{sec:lolex}
\subsection{Introduction}
Calculating the curvature $\kappa$ of the interface between two phases is
important, since it appears in the Young-Laplace formula for the capillary
pressure, $\Delta p = \sigma \kappa$.  Its value is used in \eg the Ghost
Fluid Method (GFM) (\cref{eq:jump2}), or other methods of enforcing the
jump conditions.  The normal vectors to the interface are also important,
\eg when advecting the level-set function and when reinitializing it.
Calculating these geometric quantities is straightforward in theory, using
\cref{eq:normal} and \cref{eq:curvature} to compute them from the level-set
function.

\begin{figure}[bp]
 \centering
 \subfigure[Droplets in near contact]{
   \includegraphics{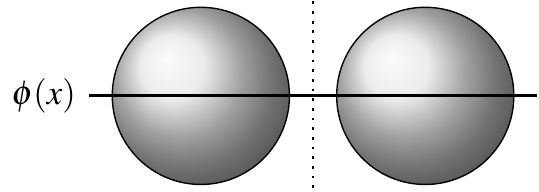}
 \label{fig:1a}}
 \subfigure[A slice of the level-set function]{
   \includegraphics{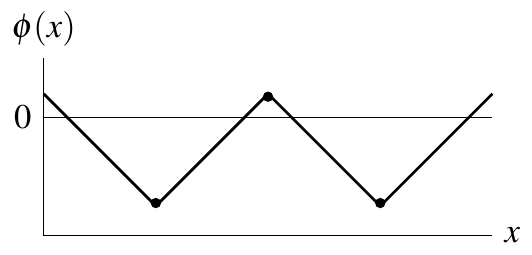}
 \label{fig:1b}}
 \caption[Level-set function of droplets in near contact]{(a) Two droplets in
 near contact. The dotted line marks a region
 where the derivative of the level-set function is not defined. (b)
 A one-dimensional slice of the level-set function.  The dots mark
 points where the derivative of $\phi$ is not defined.}
 \label{fig:curvature-problem}
\end{figure}

However, as is often the case, in practice it is not so
straightforward. The problems arise when the distance between two
interfaces is of the order $\Delta x$. This is illustrated in
\cref{fig:curvature-problem}. The derivatives of $\phi$ are not defined at
the kinks. As a result of this, the numerical stencils approximating the
derivatives of $\phi$ will often produce large, erroneous values.  When
this happens, the curvatures and normal vectors will be erroneous. For the
curvature, this error is of order $\mathcal{O}(1/\Delta x)$, which can be
several orders of magnitude larger than the correct curvature value. It
should be stressed that additional grid refinement does not solve this
problem; \eg for the simulation of colliding drops, one would have to
continue refining the grid \emph{ad infinitum}.

The earliest non-smearing approach to this problem, by Macklin and Lowengrub
\cite{macklin}, uses a modification of the directional differences for points
close to kinks, along with a mesh refinement for these points. The same authors
introduced a curve-fitting method instead in \cite{macklin06}, which is said to
be an improvement on the directional differences and a simplification. The
latter version will be referred to as the MLM (Macklin and Lowengrub Method).
Further improvements to this method, and adaptations to an on-grid framework
(\ie calculating the curvature at the grid points, not at the interface), have
been developed by Lervåg \cite{lervag-curv-conf},\cite{lervag-curv}. These
methods give good results in 2D, but are difficult to extend to 3D simulations
due to the use of curve-fitting.

An alternative approach to the problem is due to Salac and Lu \cite{salac}, and
will be referred to as the Salac and Lu Method (SLM). This approach extracts
separate bodies represented by the level-set function into their own, separate
distance functions. Only the negative parts of the level-set function are
extracted, the positive parts are reconstructed through reinitialization. This
procedure removes all kinks that are caused by two or more bodies that are
close to each other.  For a review and
comparison of the SLM, MLM and the method by Lervåg, see Lervåg and Ervik
\cite{enumath}. It should also be noted that the recent article by Focke and
Bothe \cite{focke2012} discusses a similar issue, in the context of thin
lamellae which form when liquid drops collide off-center. The authors
introduce a method which resembles the SLM, but which also has the ability to
add small amounts of liquid to the lamella region, preventing a numerical
rupture.

The method considered here is a further development of the SLM.
It is referred to as the local level-set extraction method, or LOLEX method in
short. The reason why the SLM is insufficient in some cases, as
well as the details of the present method, is given below. Suffice it to say at
this point that the present method is more general, so it applies both to the
cases considered by Salac and Lu and those considered by Focke and Bothe (except
the stabilization of thin lamellae which the latter introduce).

Another recently presented approach is due to Trontin \etal \cite{trontin}, who
consider a hybrid particle/level-set method. Their approach is to use the
information from the tracking particles to calculate the curvature and normal
vectors, with good results. This can obviously not be applied to a pure
level-set method as discussed here, or \eg a coupled level-set/VOF (CLSVOF) 
method as has recently become popular \cite{hybridlsmvof}.

The previously mentioned work by Shin \etal \cite{shin11} which introduces
a hybrid front-tracking/level-set method is another interesting approach. The
ability of their method to conserve mass globally as well as locally is
impressive, and the handling of thin filaments is better than the method
proposed in this paper. As with the approach due to Trontin et al., this method
cannot be applied to a pure level-set framework, and integrating it into an
existing level-set based code would be arduous. In comparison, the method
proposed in this paper can be implemented into a level-set framework with less
than 500 lines of code.

An approach which has not been considered here, or by other authors in the
context of level-set methods as far as we are aware, is the use of filtering.
Vliet and Verbeek \cite{vlietverbeek} study the estimation of curvature from
a discretely sampled greyscale image, using derivative-of-Gaussian filters, and
note that this outperforms a traditional curvature estimate analogous to
\cref{eq:curv-disc}.

The idea of Salac and Lu, on which the present method is based, is simple when
compared to the curve-fitting scheme used by Macklin and Lowengrub
\cite{macklin} and later by Lervåg \cite{lervag-curv-conf,lervag-curv}.
This simplicity is more in keeping with the ``spirit'' of the level-set method:
the LSM is an implicit alternative to front-tracking methods that employ curve
fitting, and this implicitness makes extending to higher dimensions
straightforward. In the same fashion, the SLM is easily extensible to 3D, while
the methods employing curve fitting are not. There are, however, some drawbacks
to the Salac and Lu method as well.

\begin{figure}[htp]
  \centering
  \includegraphics[width=0.45\linewidth]{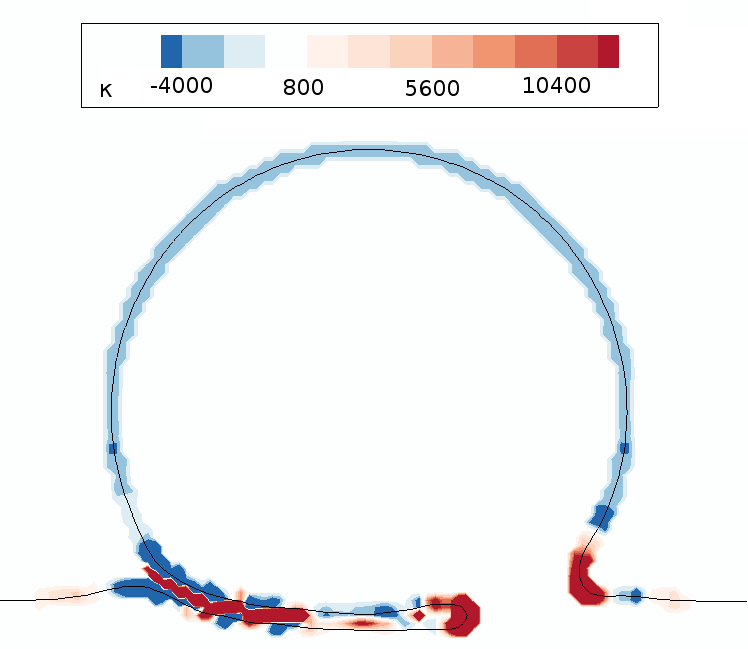}
  \caption[Incorrect curvature field from Salac and Lu method]{The curvature
    field plotted for the SLM.  Note the red curvature field inside the air
    finger between the drop and the pool, which is incorrect.  The color should
    be light blue in this area.(Figure best viewed in color.)}
  \label{fig:salac-problem}
\end{figure}

The primary issue stems from the fact that the Salac and Lu method is aware of
the global topology of the interface. A problematic area, with a kink in the
level-set function close to $\phi=0$, can be caused either by two bodies in
close proximity or by a single body folding back onto itself. In the latter
case, as illustrated in \cref{fig:salac-problem}, the Salac and Lu method falls
back to the standard discretization, and the calculated curvature will be
erroneous.  This may seem like an edge case not worth considering, but
simulations have shown that this often happens, \eg when a falling droplet
merges into a pool. As pointed out by Smereka \cite{smereka}, errors like these
can be the main contribution to unphysical area loss in a simulation.
Another situation where this would often be the case is in tumor simulations
like those performed by Macklin and Lowengrub, as can be seen in \eg
\cite[Figure~6]{macklin}.

\subsection{The idea of the LOLEX method}
\label{lolex}

The method presented here tries to combine the best of the SLM with the best of
the MLM. As illustrated in the
previous section, the SLM is aware of the global topology of
the interface, which is problematic in some cases. The MLM
does not have this problem, as its curve fitting considers only the local
area, but as previously stated it does not extend easily to 3D. A natural
workaround to the ``global awareness'' is to make the Salac and Lu method
consider only the local topology; say, a 10\x10\x10 cube around the point
where we calculate the curvature.

In the following, we assume the level-set function to be located on a uniform
mesh on a single CPU.
The proposed method can be adapted in a straight-forward manner both 
into a domain-decomposition and a mesh-refinement framework. We do not discuss
this in further detail here.

Since the SLM relies on reinitialization to remove kinks,
a potential problem with this approach is computational efficiency, as
reinitialization can be time-consuming.
 To avoid this problem, we want to use the standard discretization as
much as possible, only resorting to the LOLEX method when we have to, \ie when
kinks in $\phi$ are close to the interface.
To easily identify kinks, we use the quality function $Q(\v{x})$ which was
introduced by Macklin and Lowengrub in \cite{macklin}. It is defined as
\begin{equation}
  Q(\v{x}) = \|1 - \grad\phi(\v{x})\|_2,
  \label{eq:quality}
\end{equation}
\ie the deviation of $\phi$ from a signed distance function, measured with the
2 norm. If
$\rm{max}(Q(\v{x}_{i,j,k})) > \eta$ for $\v{x}_{i,j,k}$ in a 3\x3\x3 cube around
the current grid point, we use the LOLEX method.
A value of $\eta = 0.005$ is used here, and is seen to perform
well. That is, the number of grid points where the LOLEX method is used becomes small
compared to the total number of grid points. This keeps the computational cost
low. The effect of varying $\eta$ can be seen in \cref{fig:curv-inf-norms}, as
discussed in \cref{sec:drop-fall-test}

To further decrease the computational cost, we use the ``narrow band''
level-set method introduced in \cite{tubes}. This means that quantities such as
the curvature are only calculated in a narrow band around the zero level set,
where they are needed.

Having briefly presented the idea behind the present method and the scope in
which it will be used, we give here a step-by-step outline of it, see
\cref{fig:algorithm}. 2D notation is
used for clarity, but all steps are easily extensible to 3D. In this
outline, a few arrays are introduced for storing data: \texttt{lookphi} is
a copy of the global $\phi$ for the local area we are considering,
\texttt{bodies} indicates the bodies present using increasing integers, and
\texttt{locphi} holds the local $\phi$s that are extracted from the global
$\phi$ and then refined into more accurate representations of the local bodies
present. The quantities \texttt{ilmax}, \texttt{jlmax} and \texttt{klmax}
represent the number of grid points, in the $x$, $y$ and $z$ directions
respectively, of the \emph{local} grid. The values of \texttt{ilmax},
\texttt{jlmax}, \texttt{klmax} are all set to 7 in the simulations performed
here. Their values are independent of the global grid size. Sensible values of
these are between 5 and 11; since they must be odd, smaller than 5 gives too
low resolution, and larger than 11 starts eliminating the advantage of using
a localized method. The value of 7 used here gives good results, and increasing
it to 9 gives only a small change while increasing the computational cost. 
In the limit \texttt{ilmax} $\to$ \texttt{imax} etc.\ the method of Salac and Lu is 
recovered.

\begin{figure}
  \begin{itemize}
    \renewcommand{\labelitemi}{$\hookrightarrow$}
    \renewcommand{\labelitemii}{$\hookrightarrow$}
    \renewcommand{\labelitemiii}{$\v{-}$}
    \item Loop over the computational domain using indices \texttt{i,j}.
    \item If ($\v{x}_{i,j}$ not close to interface) do nothing. A point is
      defined as close to an interface if all $\phi(\v{x}_{n,m})$ for $(n,m)
      \in [i-1,i+1]\times[j-1,j+1]$ is either negative or positive.
    \item Else if ( $Q(\v{x}_{n,m}) \le \eta \; \forall (n,m) \in
      [i-1,i+1]\times[j-1,j+1]$ ) use ordinary method.
    \item Else use LOLEX method:
      \begin{itemize}
        \item Copy $\phi$ in a \texttt{[-1,ilmax+2]*[-1,jlmax+2]} square around
          \texttt{i,j} into the \texttt{lookphi} array.
        \item Identify the bodies present in the
          \texttt{[0,ilmax+1]*[0,jlmax+1]} square, store this in the
          \texttt{bodies} array.
        \item For each body, extract the relevant part of the \texttt{lookphi}
          array into \texttt{locphi(:,:,bodyno)}. This array has 3 ghost
          cells on the boundary outside \texttt{ilmax*jlmax}; these are
          not used until the extrapolation further down. Extracting means:
          \begin{itemize}
            \item copying \texttt{lookphi} for the internal points of
              \emph{this} body
            \item copying \texttt{lookphi} for external points that are not
              next to more than one body
            \item explicitly reconstructing the signed distance for external
              points that are next to more than one body
            \item setting a value of \texttt{2*dx} for all other points
          \end{itemize}
        \item Once the \texttt{locphi} array has been filled for all bodies,
          the values are extrapolated into the ghost cells. The extrapolation
          is zeroth-order, as will be explained further down.
        \item The \texttt{locphi} array is then reinitialized for all bodies.
          This erases the problematic kink, as well as the value of
          \texttt{2*dx} which was set previously. Thus this value is
          unimportant, as long as it is $> 0$.
        \item Using these local $\phi$'s, the curvature and normal vectors
          can be calculated for each body. The curvature and normal vectors
          corresponding to the body which is closest to the current grid
          point are used.
      \end{itemize}
  \end{itemize}
  \caption{A step-by-step outline of the LOLEX algorithm.}
  \label{fig:algorithm}
\end{figure}

The steps in the algorithm that warrant further comments are:
identifying the bodies present, explicitly reconstructing the signed distance,
extrapolating to the ghost cells, and reinitializing. These will be considered
further in the next section and subsections.

\subsection{Details of the method}
\label{lolex-details}
Some steps of the algorithm outlined need further
explanations. This is either because they are too technical to be fully
described in the previous short outline, or because they have not been properly
motivated yet. The steps that will be considered are identifying the bodies
present (\cref{bodyscan}), explicitly reconstructing the signed distance
(\cref{reconstruct}), extrapolating to the ghost cells (\cref{extrapolation}),
and reinitializing (\cref{reinitialization}).

\subsubsection{Identifying the bodies present}
\label{bodyscan}

To identify the bodies present, a  recursive routine  is used,
which starts  at  a seed  point in  a body  and iterates through  the entire 
body,  marking it  as a body in  the \texttt{bodies}
array. The seed point is found by scanning the computational domain for
points with $\phi <0$. The recursive routine is called \texttt{bodyscan} here.
The \texttt{bodies} array  starts with  a value  of
\texttt{unchecked},  and bodies found are marked using increasing integers.
The recursive subroutine will have  marked
the entire first body when its first call returns.

A final point to note about the routine given here is that even though
a recursive subroutine is used, memory usage will not be problematic. This is
because the routine operates on a small array whose size is independent of the
grid size. In 3 dimensions and with the presently used size of the local area,
the array \texttt{bodyscan} would have \texttt{11*11*11} $=1331$ elements.  This
routine can maximally be called $1331$ times, giving a worst-case memory
consumption of 13.5 MB. In reality this number would typically be less than half
of that. This will not cause memory problems, although it is too
large to fit in the CPU cache for some processors. The performance impact has
not been tested here, as the 3D calculations are only considered as
a proof-of-concept, and have not been optimized for speed. In 2D the memory use
is naturally much smaller.


\subsubsection{Explicit reconstruction of the signed distance}
\label{reconstruct}
For some points with $\phi > 0$, two or more bodies are within $\Delta x$ of the
point. This means that the value of $\phi$ is probably incorrect, since it has
to be the distance to two separate bodies at the same time.  We will call such
points ``dependent points''. These points are found using the \texttt{bodies}
array: if this array has more than one unique positive integer value in the four points
adjacent to the present point, it is dependent. Because $\phi$ is likely
incorrect for dependent points, we discard its value, and instead explicitly
reconstruct the distance to the relevant interface. The procedure used is due to
Adalsteinsson et al.\ \cite{adalsteinsson}.

When we consider such a dependent point, it lies right next
to two interfaces. When reconstructing the distance, only one interface is
of interest, so the other one is momentarily removed. Note that the signed
distance is always positive for exterior points, so it is just the normal
distance.

\begin{figure}
  \centering
  \subfigure[]{
    \includegraphics{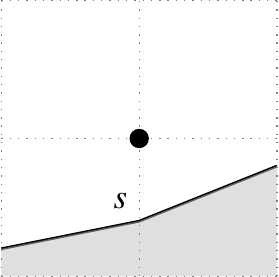}
    \label{fig:neighborhood-points-a}}
  \hspace{1em}
  \subfigure[]{
    \includegraphics{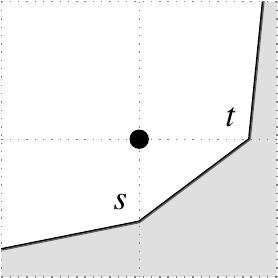}
    \label{fig:neighborhood-points-b}}\\
  \subfigure[]{
    \includegraphics{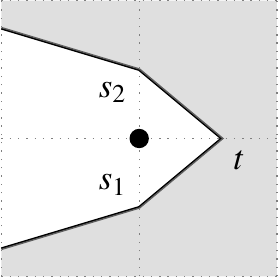}
    \label{fig:neighborhood-points-c}}
  \hspace{1em}
  \subfigure[]{
    \includegraphics{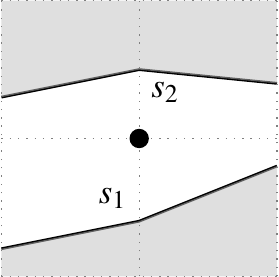}
    \label{fig:neighborhood-points-d}}
  \caption{Cases for the neighborhood of a point.}
  \label{fig:neighborhood-points}
\end{figure}

The procedure in \cite{adalsteinsson} is as follows. The point $(i,j)$ which we
are considering is next to the interface of current interest. We ignore all
other interfaces.  Up to rotational symmetry, there are four possible cases.
An illustration of these cases can be seen in \cref{fig:neighborhood-points}.

We examine the four cases (\textbf{a} to \textbf{d}) more closely:
\begin{description}
  \item[a] The interface crosses one of the lines from $(i,j)$ to its four
    neighbors. In this case, we use the distance to the interface
    along this line as our distance. This distance is given by
    \begin{equation}
      s =  \Delta y + \phi(i,j-1)
      \label{line-distance}
    \end{equation}
    where we have assumed that $(i,j-1)$ is the neighbor on the other side of
    the interface. Since this neighbor is an internal point, it has $\phi
    < 0$. The distance to the interface is the distance to the neighboring
    grid point ($\Delta y$) minus the distance from that grid point to the
    interface, which gives this formula. It is best to use only the
    $\phi$-value inside the body, since it is less likely to be distorted.
  \item[b] The interface crosses two of the lines, and these two lines make out
    a corner of the 2\x2 grid around $(i,j)$. In this case we use the shortest
    distance to the straight line between the two points of intersection. The
    distance $d$ is given by the formula
    \begin{equation}
      \left( \frac{d}{s} \right)^2 + \left( \frac{d}{t} \right)^2
      = 1 \rm{.}
      \label{eq:corner-distance}
    \end{equation}
    As long as $s^2+t^2 \ne 0$ this equation can be solved, and the positive
    solution is
    \begin{equation}
      d = \frac{st}{\sqrt{s^2+t^2}}\quad\rm{.}
      \label{eq:corner-distance-solved}
    \end{equation}
    If we have $s^2+t^2 = 0$, then $s=t=0$, so it is obvious that the distance
    to the interface is $d = 0$.
  \item[c] The interface crosses three lines. We construct the two straight
    lines between the points of intersection, and use the shortest distance to
    either of these two lines, given by
    \begin{equation}
      \left( \frac{d}{\min(s1,s2)} \right)^2 + \left( \frac{d}{t}
      \right)^2 = 1 \rm{.}
      \label{eq:bulge-distance}
    \end{equation}
  \item[d] The interface crosses two lines. These lines are on opposite sides
    of the point $(i,j)$. In this case, we use the shortest of the two
    distances, so $d = \min(s1,s2)$.
\end{description}

These formulae can be extended to three dimensions, where the possible cases are
more numerous. In 3D, the central point has two additional neighbors. This
means there are more variations in addition to the cases considered above. This
is not considered in detail here.


\subsubsection{Extrapolation}
\label{extrapolation}
\begin{figure}[tbp]
        \begin{center}
                \subfigure[Zoom in on global level set]
                {\includegraphics[width=0.35\textwidth]
        {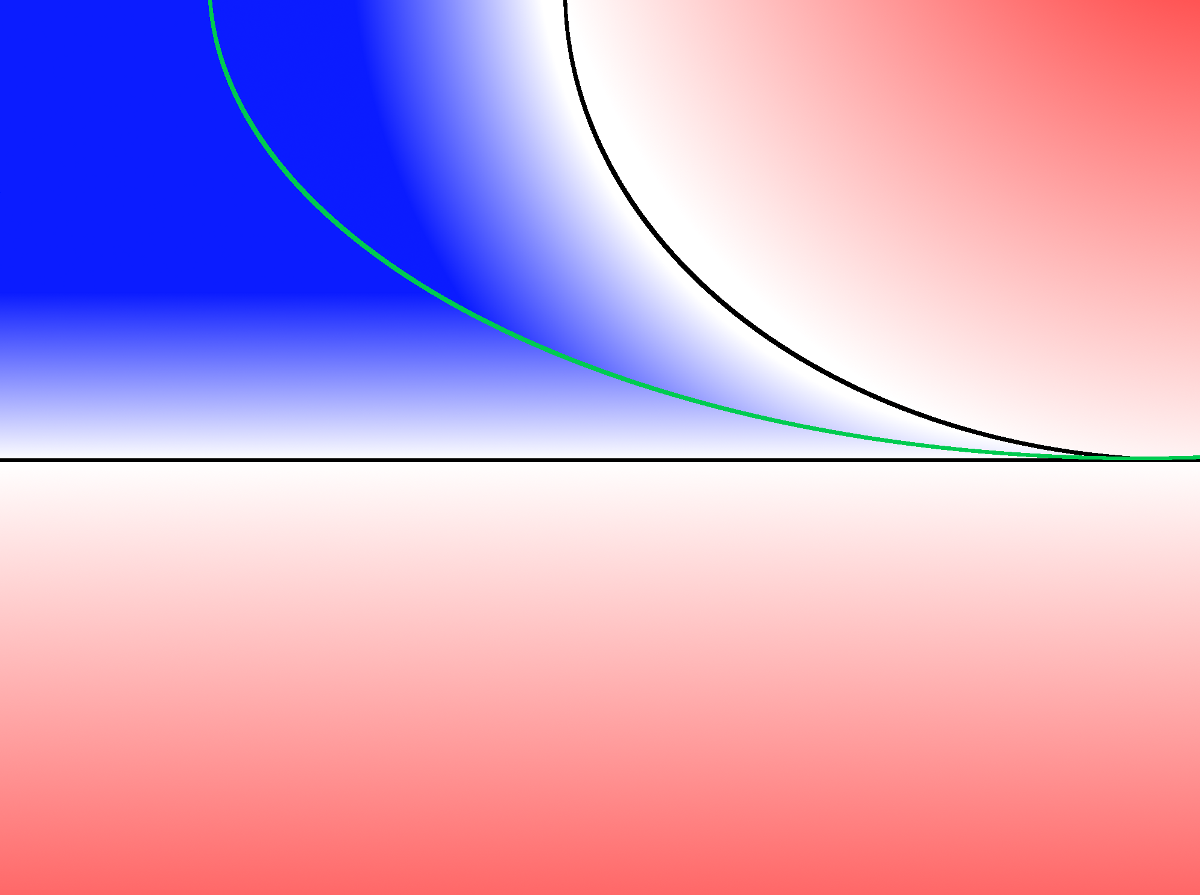}}
    $\quad$
                \subfigure[Extracted local level set]
                {\includegraphics[width=0.35\textwidth]
      {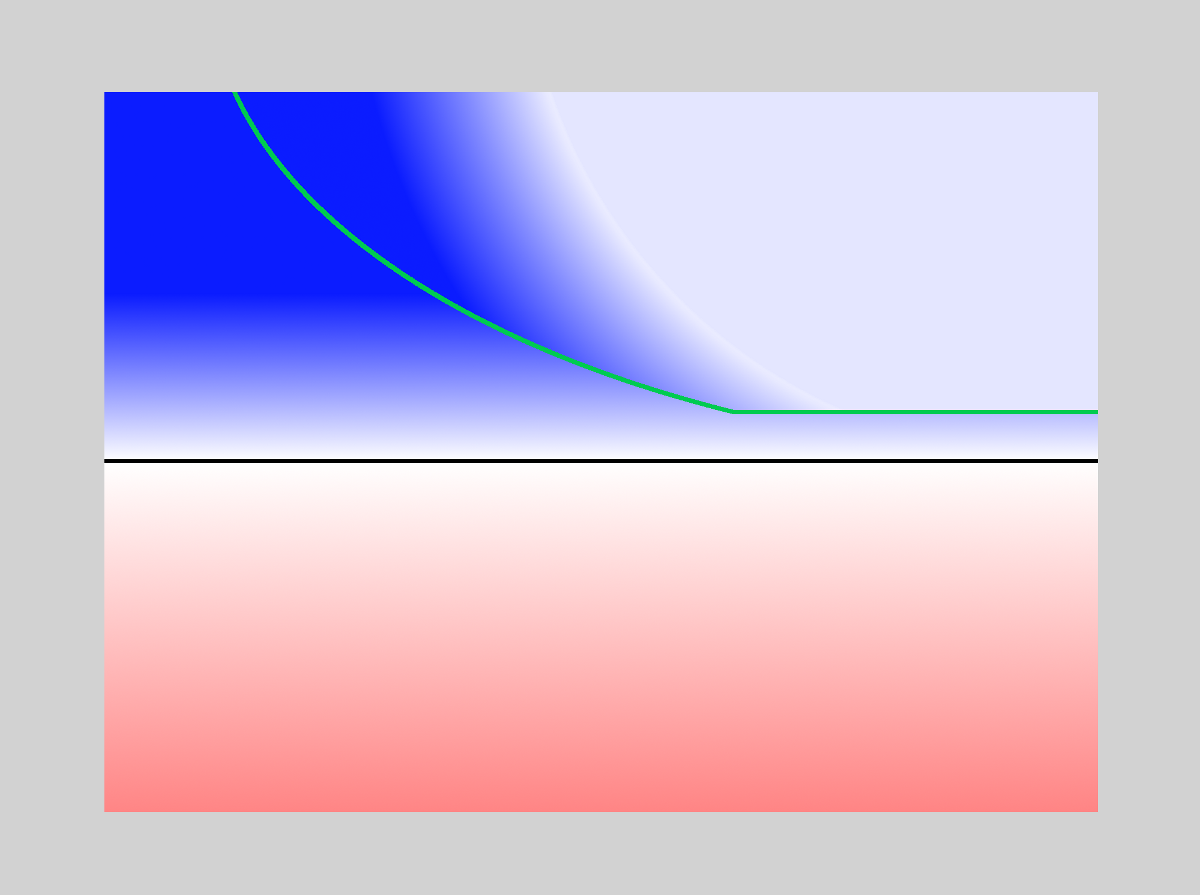}}
    \\
                \subfigure[First order extrapolated]
                {\includegraphics[width=0.35\textwidth]
      {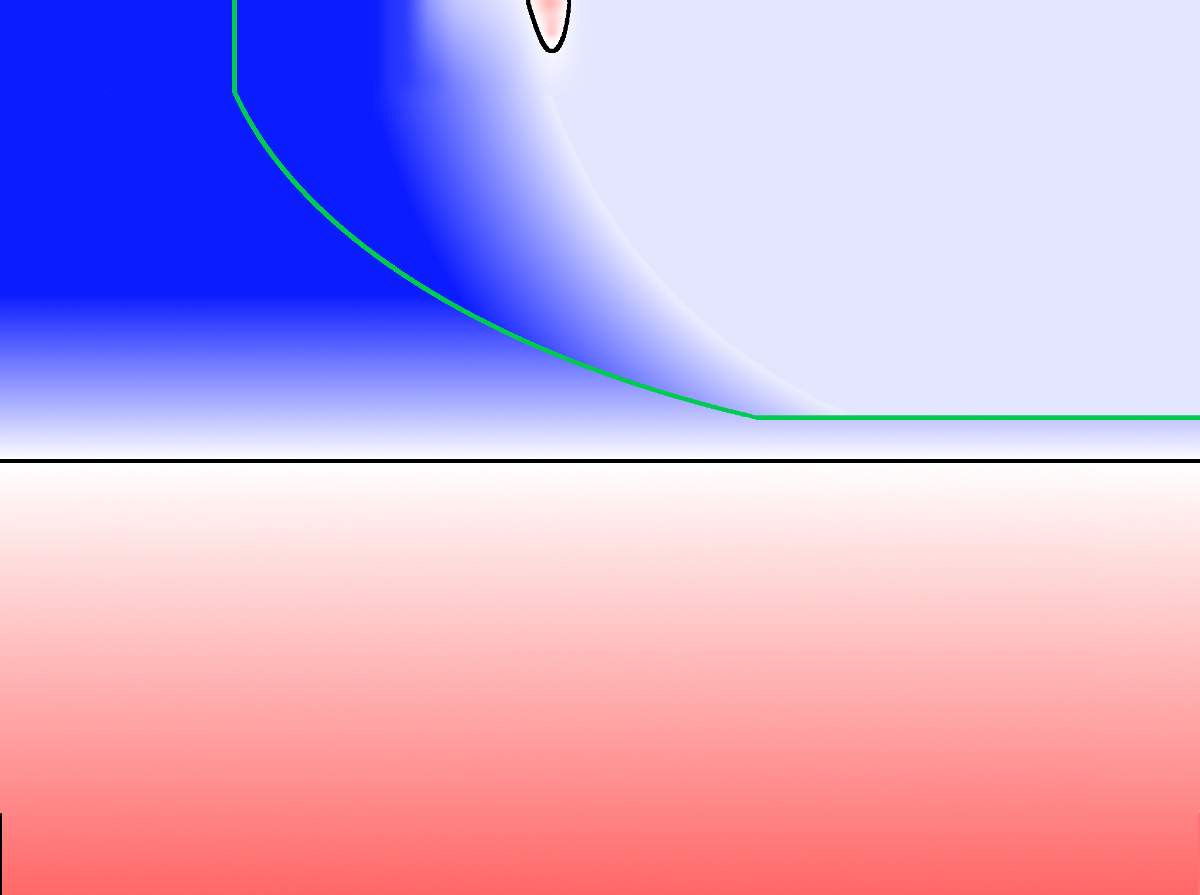}}
    $\quad$
                \subfigure[Zeroth order extrapolated]
                {\includegraphics[width=0.35\textwidth]
      {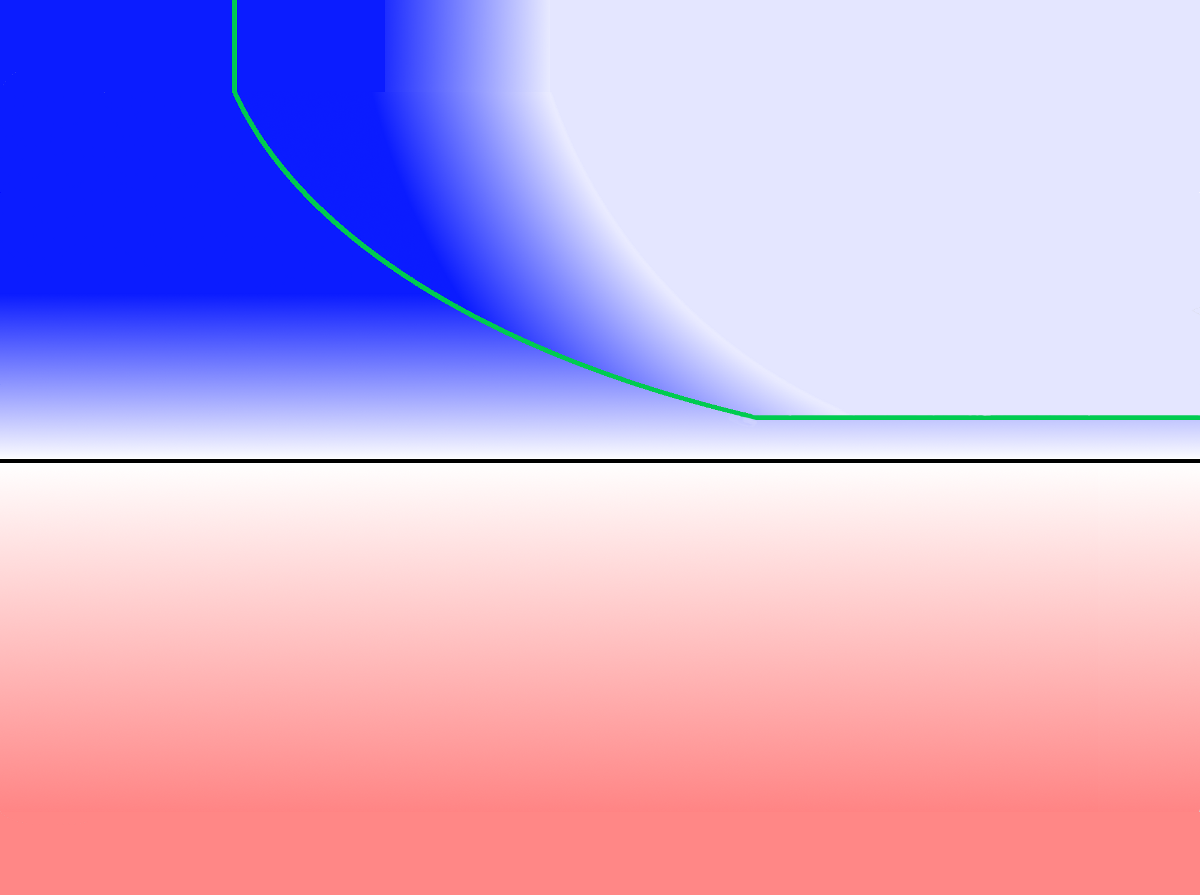}}
    \\
                \subfigure[First order, reinitialized]
                {\includegraphics[width=0.35\textwidth]
      {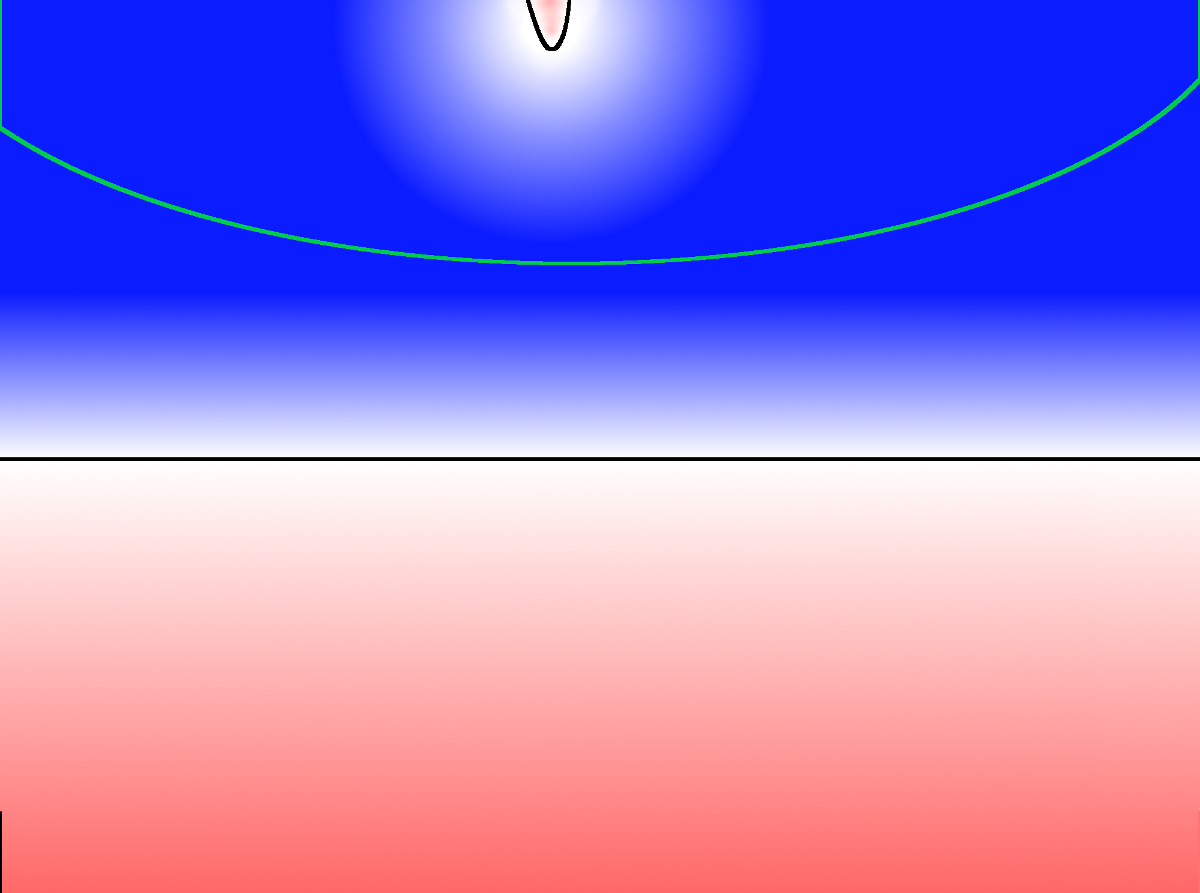}}
    $\quad$
                \subfigure[Zeroth order, reinitialized]
                {\includegraphics[width=0.35\textwidth]
      {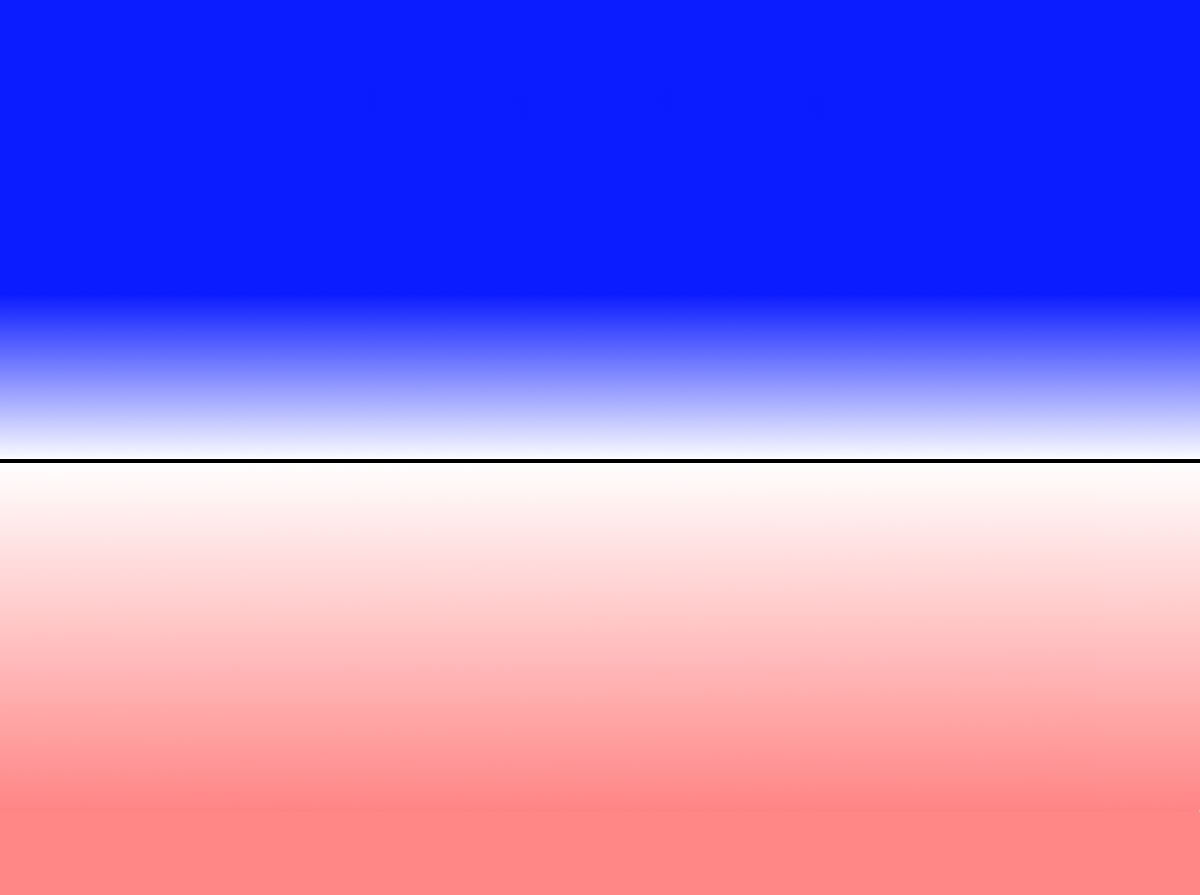}}
  \end{center}
  \caption[Extraction, extrapolation and reinitialization of the local level
  set]{Extraction, extrapolation and reinitialization of the local level set
    is shown, for the lower body in Figure (a). Red indicates a negative value, blue a positive value,
    and white indicates zero. The green lines indicate kinks in the level set
    function, and the black lines are the zero level sets. A detailed
    explanation of the figures is given in \cref{extrapolation}. (Figure best
  viewed in color.)}
\label{fig:extrap-demo}
\end{figure}

After the interior of the \texttt{locphi} array has been filled, the ghost cells
must be filled before we can reinitialize the local $\phi$. Two ways
of doing this are illustrated in \cref{fig:extrap-demo}. A first approach is
to use linear extrapolation, which should work well since $\phi$ is a linear
function in 1D. However, it turns out that this does not work.
A fundamental property of the reinitialization equation (\ref{eq:reinit})
is that its characteristics originate at the interface $\phi=0$. This is why the
present method (and the SLM) works -- we only need a few cells
directly next to the interface to have the correct value of $\phi$, and
reinitialization will fix the rest. It also means that reinitialization will
never move the position of the interface, which is a desirable property in
general.

The problem with linear extrapolation occurs when we extrapolate starting on the
opposite side of the kink from the interface. In this case, the values of the
local $\phi$ are tending towards $0$ from above, which means that extrapolation
can reintroduce the other body (which we removed in the first place). When this
happens, reinitialization cannot fix the values beyond the kink, since it cannot
move the interface reintroduced by extrapolation. A straightforward alternative
is to use a zeroth-order extrapolation. This means simply copying the values
along the edges into the ghost cells. It is obvious that this will never cross
$\phi=0$, so reinitialization works as intended. 

The difference between these two is shown in \cref{fig:extrap-demo}. In (a),
a zoom in on the global level set of a droplet touching a pool is shown. In (b),
the local level set of the lower body (the pool) is shown after extraction and
explicit reconstruction. Here, the values on the edges are not set, indicated in
grey.  In (c), the same is shown after first-order extrapolation, and in (d)
after zeroth-order extrapolation. In (e), the first-order extrapolated $\phi$ is
shown reinitialized, and in (f) the zeroth-order extrapolated $\phi$ is shown
reinitialized. Note in particular that in (e), a kink still exists after the
entire procedure (green line), so the geometric quantities calculated would
still be wrong if the derivatives cross the kink.   

The corner cells on the boundaries must also be set. Here, these all get the
value from the corresponding corner of the internal grid.


\subsubsection{Reinitialization}
\label{reinitialization}
When the extracted local level-set has been extrapolated, it must then be
reinitialized before the geometric quantities are calculated. This is essential
in order to have good values of the level-set function outside the interface.
The entire LOLEX method hinges on the fact that reinitialization restores the
local level-set to a signed distance function, so that ordinary discretizations
will not give errors. This is not entirely straightforward, however.

When reinitializing, we require at least some points on either side of the
interface with decent $\phi$-values, \ie $\phi$ being the signed distance to the
interface. In addition to this, we need to know the
smeared sign function, and most crucially, the normal vectors at the
interface. Thus we are faced with a bootstrapping problem: accurate normal vectors
are required in order to accurately calculate the normal vectors. This is
only a problem when the global interfaces are very close; when there
is a moderate distance (\ie more than one grid point between the interfaces),
the normal vectors can be calculated at the interface using the local level-set.

The solution to this conundrum is to exploit the redundant information which is
stored in the level-set function. To illustrate this redundancy, imagine that
you are walking along a normal vector to the interface. At each grid point you
pass, you are told the current distance to the interface. As long as you do not
pass any kinks, this information is redundant: using the value at the first grid
point you pass, you can calculate the value at the next grid point, and the one
after that, given that you know the grid spacing.

\begin{figure}[tbp]
  \begin{center}
    \includegraphics[width=0.8\linewidth]{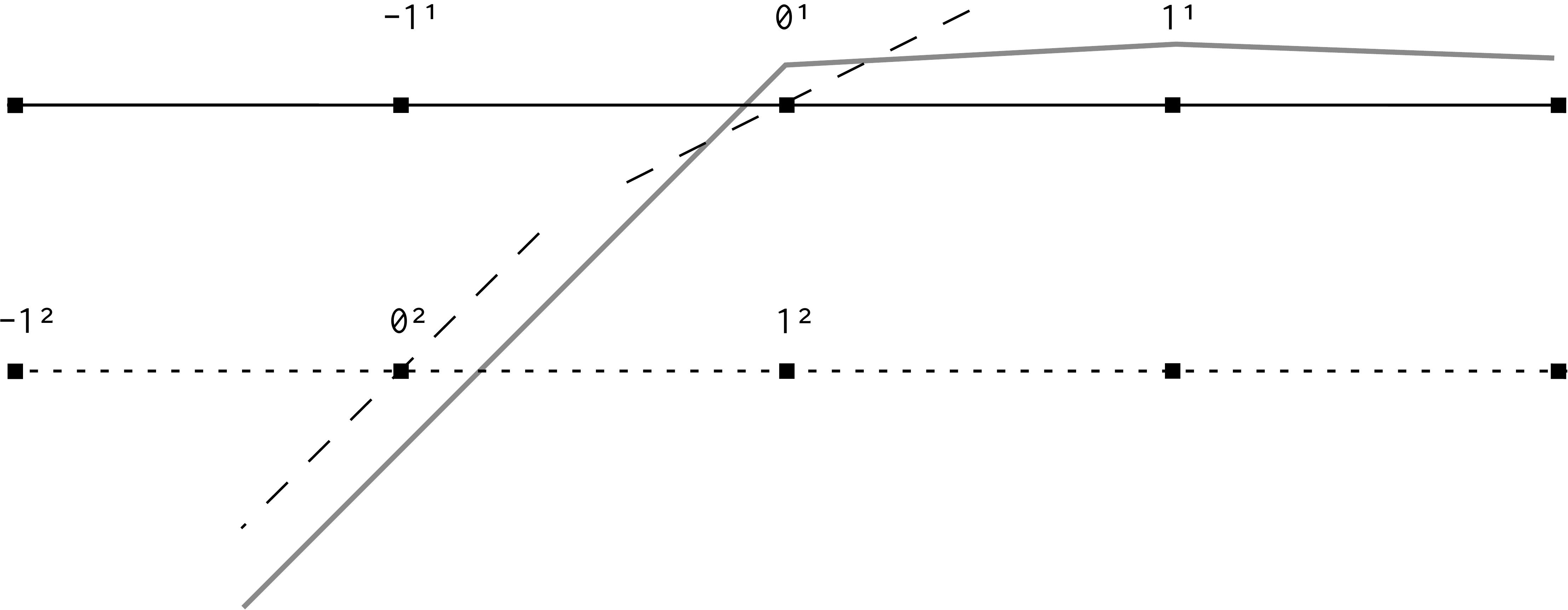}
  \end{center}
  \caption{Why we reinitialize from a lower level set: At the lower level set,
  indicated by the dotted line, values of \eg $\grad\phi$
  are more accurate at the grid point which is closest to the
  grey line than for the zero level set. The grey line indicates the local level-set
  function $\phi$. The dashed lines indicate $\grad\phi$ calculated using
  central differences.}
  \label{fig:1D-local-reinit}
\end{figure}

What this means for the present case is that we have information inside the
current body that we can use. Most importantly, we can calculate the normal
vectors without problem for internal points. This means that we can reinitialize
a level set different from $\phi=0$, \eg $\phi=-0.8\Delta x$, and get
essentially the correct $\phi$ afterwards. We are not guaranteed to get exactly
the correct $\phi$, but as we cannot obtain the correct $\phi$ anyway, we
will settle for a good approximation. An illustration of this in 1D is shown in
\cref{fig:1D-local-reinit}, where the extracted local level-set function $\phi$
is shown in grey. Note that \eg the value of $\grad\phi$ at the grid point
\texttt{0$^2$}, shown with a dashed line, is much closer to 1 than the value at
the grid point \texttt{0$^1$}. When the lower level
set is used, we momentarily move the interface further to the left in this
figure, so the grid point \texttt{0$^2$} is closest to the interface. It is
obvious that we have a better chance of restoring a signed distance function
with the correct location of the interface if we reinitialize from the lower
level set.

The value of $-0.8\Delta x$ used here gives the most accurate results. If the
value is too close to zero, the benefit of reinitializing from a lower level
set is reduced. However, if the value is too large, we risk having this lower
level set too close to the edges of the local domain, and we increase the
potential error caused by reinitializing from a different level than zero. The
optimal result is afforded by choosing a value somewhat below $-0.5 \Delta x$,
since this ensures that the grid point \texttt{0$^2$} is always closest to the
interface, while minimizing errors from the edges of the domain.

Another problem solved by this is the fact that the values directly outside the
zero level set may be incorrect in some cases. In particular, this happens when
an outside grid cell is not flagged as dependent, but its value of $\phi$ still
deviates from that dictated by a signed distance function. Tests have shown that
this sometimes occurs, and that it distorts the reconstructed local level set.

\begin{figure}[tbp]
  \centering
  \noindent\makebox[1.2\textwidth]{\hspace{-90pt}
    \subfigure[Before]
    {\includegraphics[width=0.38\textwidth]
      {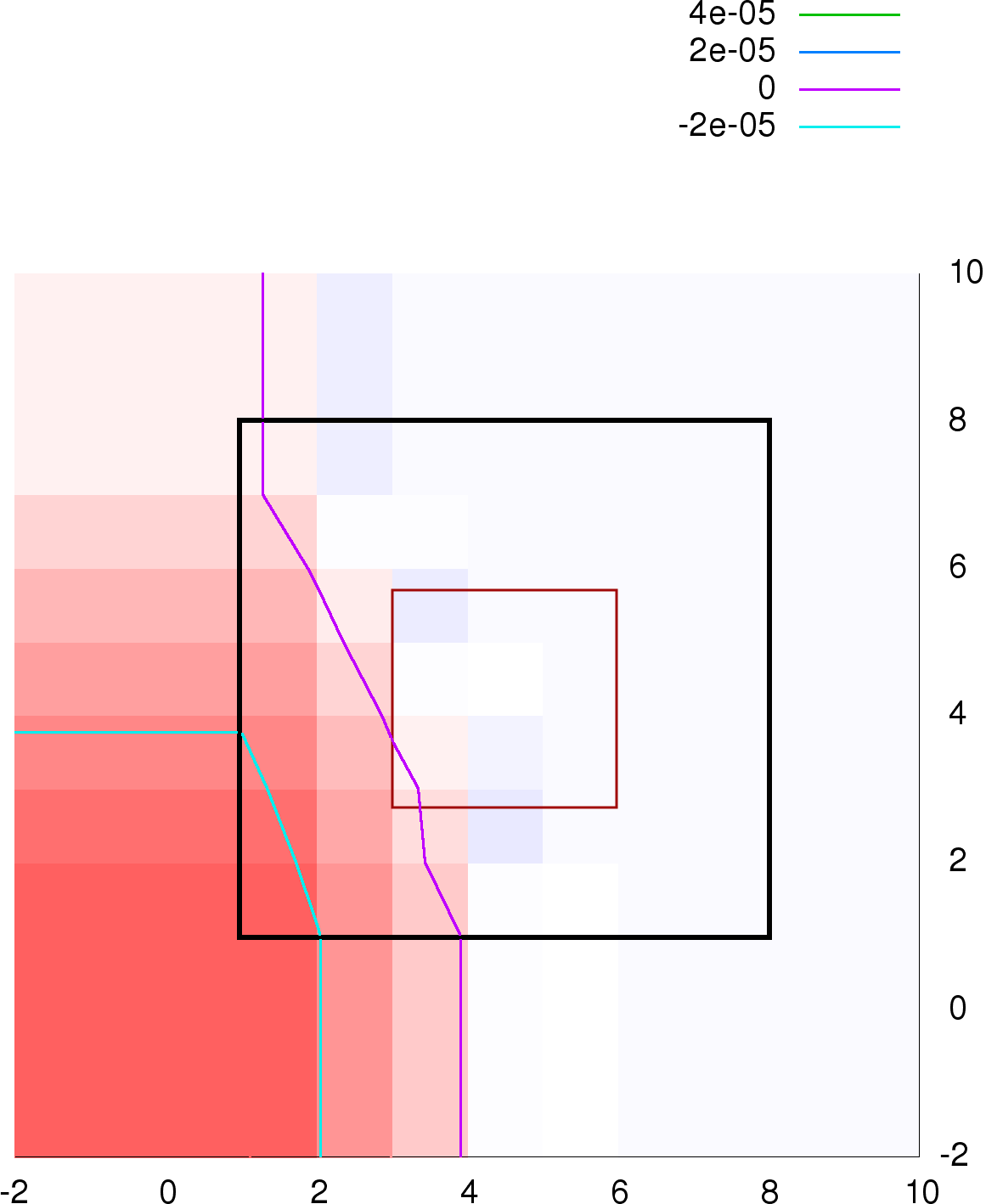}}
    $\;$
    \subfigure[From $\phi=-0.8\Delta x$]
    {\includegraphics[width=0.39\textwidth]
      {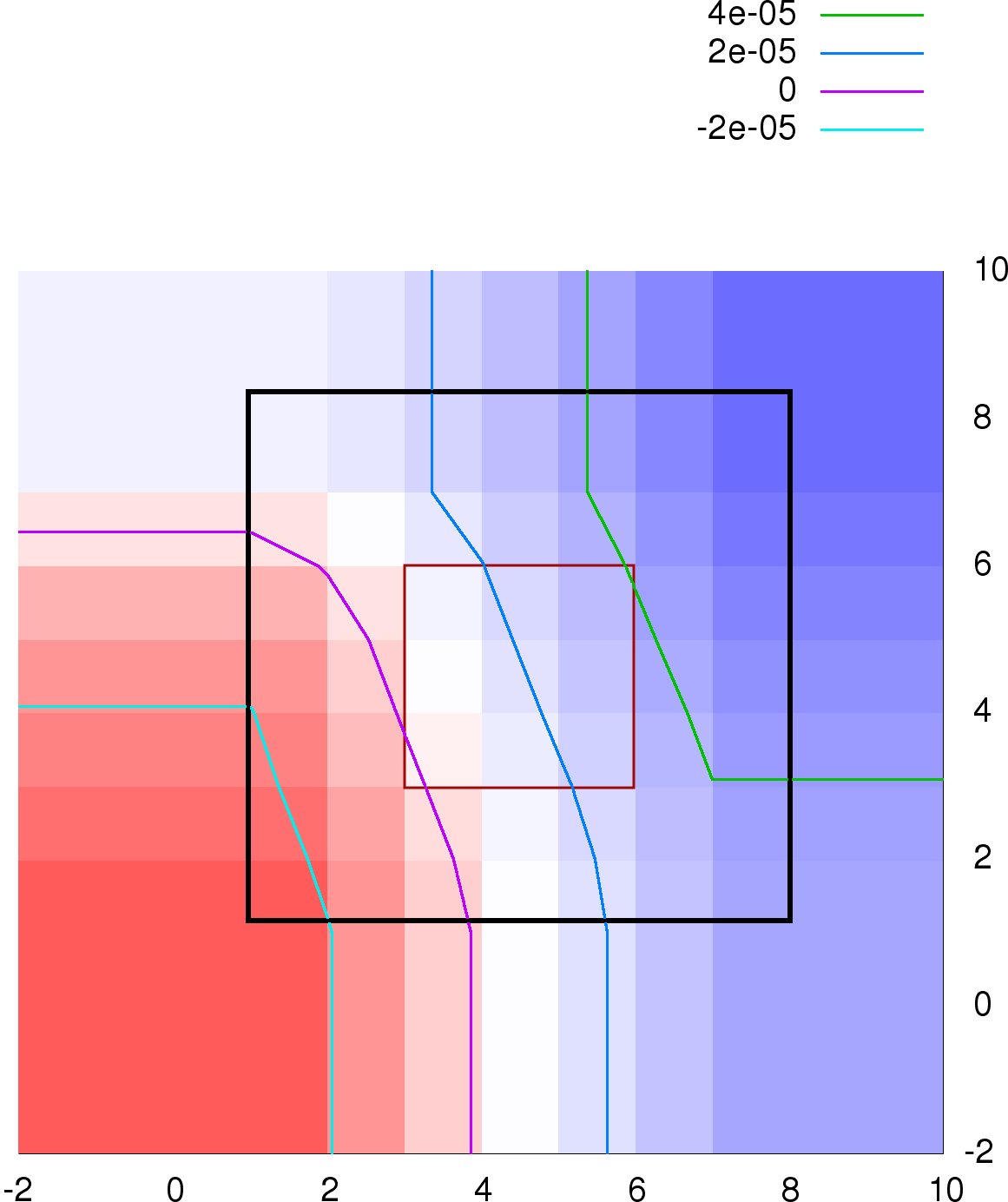}}
    $\;$
    \subfigure[From $\phi=0$]
    {\includegraphics[width=0.39\textwidth]
      {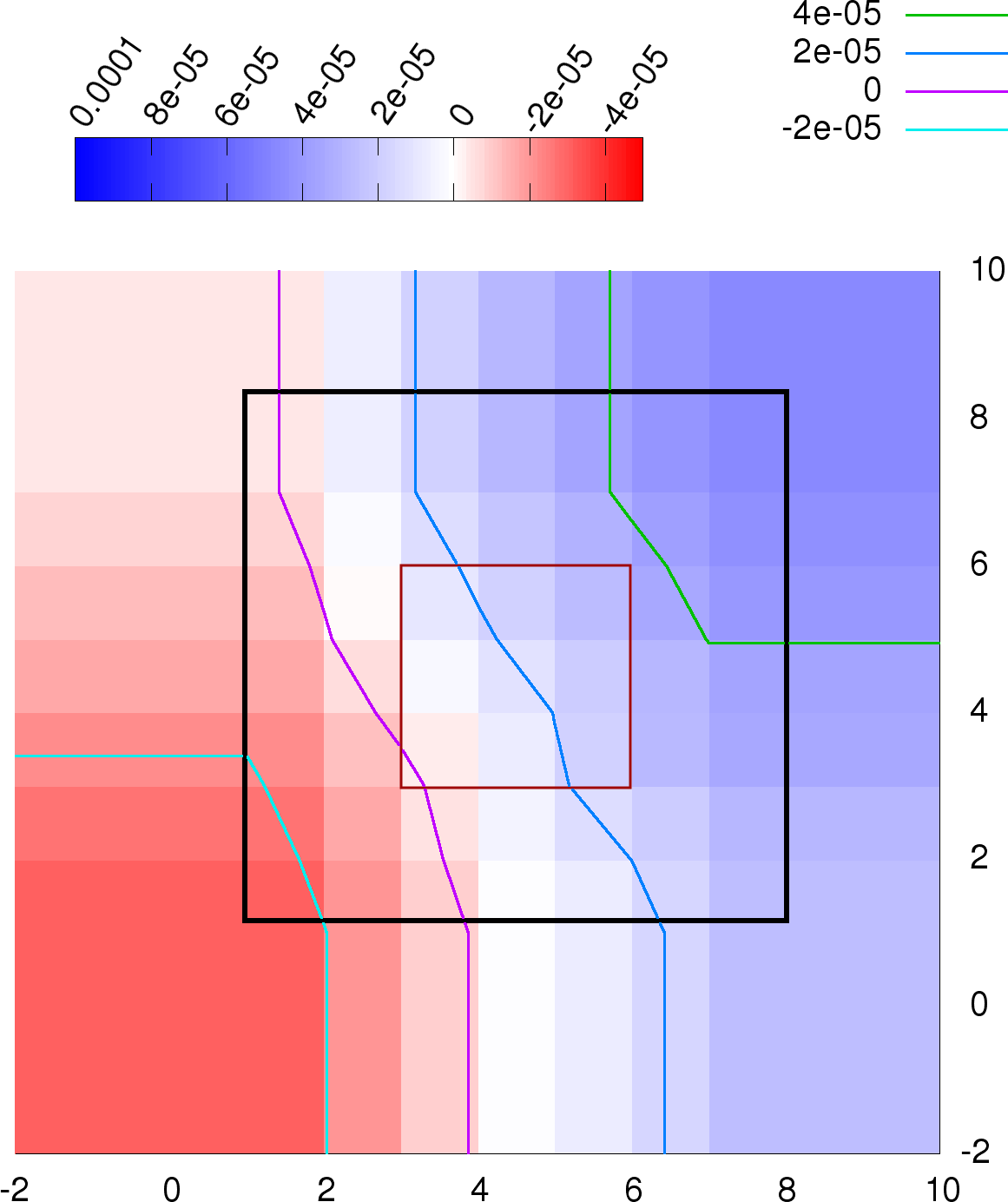}}
  }
  \caption[Reinitializing from lower local level set]{ The LOLEX method on
    a global level set which is distorted due to reinitialization of very close
    bodies. The global bodies are two concentric circles. 
    (a) Local $\phi$ before reinitialization. (b) Local $\phi$
    reinitialized from $\phi=-0.8\Delta x$. (c) Local $\phi$ reinitialized from
    $\phi = 0.0$. The black square indicates the boundary to the ghost cells,
    and the red square indicates the 3\x3 central points that are used in the
    final curvature calculation.}
  \label{fig:reinit-lower}
\end{figure}

\begin{figure}[tbp]
  \begin{center}
    \includegraphics{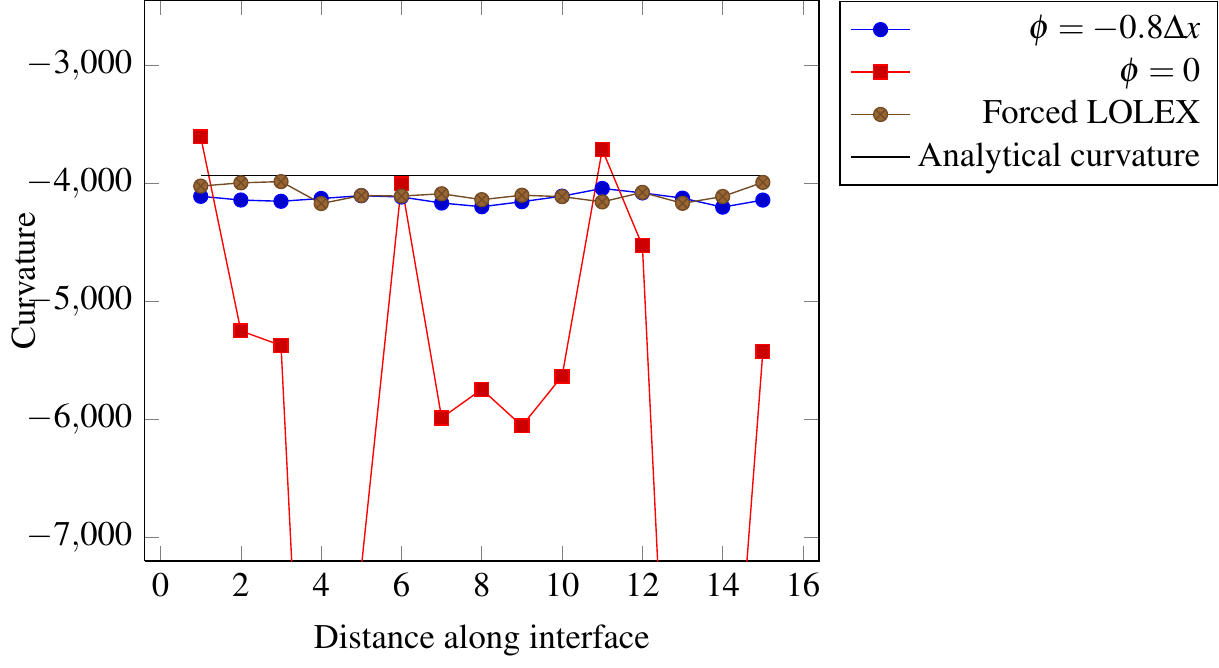}
  \end{center}
  \caption{Lineplots of the curvature along the interface when reinitializing
  from both the zero level set and a lower level set. Also shown are the
  curvature calculated when forcing use of the LOLEX method on a single
  interface, and the analytical curvature.}
  \label{fig:avg-line-exponent}
\end{figure}

Reinitializing from a different level may sound somewhat complicated to do, but
the implicit formulation springs to the rescue again. To reinitialize from
a lower level set, we simply add a positive constant to $\phi$ at every local
grid point, call the reinitialization routine on this $\phi$, and then subtract
the same constant from the reinitialized $\phi$. The effect of this is
illustrated in \cref{fig:reinit-lower}, which is an extreme case. Here,
reinitialization of two very close bodies (concentric circles) has distorted the
global level-set function close to and outside the interfaces. The reinitialized
local level-set function is also wrong, but the one which is reinitialized from
a lower level set gives a much smoother representation of the interface, which
agrees with the contour lines further into the body. This smoother
representation will, in turn, give a significantly more accurate curvature.
A plot of the curvature calculated with and without this improvement is shown in
\cref{fig:avg-line-exponent} for the concentric circles case; this global
interface configuration can also be seen in \cref{fig:normals} further down.
This plot shows the curvature along the inner circle. It is seen that the
improvement is large, particularly in this case when two interfaces are close.
The curvature calculated using the standard method is not shown, as it is
outside the $y$-axis range in this figure.

While the curvature calculated using the LOLEX method is close to the analytical
value, there is still a more or less constant error of 1--2\%. It turns out that
this error is caused by the reinitialization of the local level set, as is
indicated in this figure as well. The line captioned `Forced LOLEX' shows the
LOLEX method used on a single interface corresponding to the inner circle. Here,
the level-set function is correct and the standard method gives an error for the
single interface which is smaller than the line width in this figure. When we
force the use of the LOLEX method, the only difference from the standard method
is the extrapolation and reinitialization, meaning that these must be the
culprits. To mitigate this, a more accurate reinitialization procedure could be
used, \eg the HCR method due to Hartmann \etal \cite{hartmann10}.

\subsubsection{Parameters of the method}
In the LOLEX method as presented here, there are a number of
parameters that can be varied. An overview of these is given here, along with
the values used presently, and sensible ranges, in \cref{tab:parameters}.

\begin{table}
  \centering
\caption{Parameters used in the LOLEX method, along with values used
and sensible ranges.}
\label{tab:parameters}
\begin{tabular}[tbp]{lrc}
\toprule
Parameter & Value & Sensible range \\
\midrule
Local grid size & 7 & 5--11 \\
Gradient threshold $\eta$ & 0.005 & 0.01--0.001 \\
Reinit. level set & $-0.8\Delta x$ & $-1.0\Delta x$ to $-0.5\Delta x$ \\
\bottomrule
\end{tabular}
\end{table}

After the local level sets have been extracted correctly, the standard
discretizations can be used to calculate the normal vector and curvature.  As
the curvature and normal vector cannot be multiply defined at a single grid
point, we must combine the information from different local level sets.  To do
this, we simply select the one corresponding to the interface which is closest
to the central point.


As the present method uses reinitialization on a local grid for each grid point
where it is used, the performance impact of the method could become large.  To
avoid this, the quality function $Q(\v{x})$ is used to restrict the use of the
method.  In a typical falling drop simulation, the present method will only be
used in a small percentage of the total number of time steps, and even then, it
will typically not be used for all points along the interface.  This means the
computational cost of the present method has a low impact on the
total runtime of a simulation. We note that the computational cost is lower than
in the method of Salac and Lu, since that applies reinitialization to the
entirety of the bodies present.


\subsection{Summary}
In this section the presently used LOLEX method has been described
in detail. The method is used for grid points where the level-set function
deviates from being a signed distance function, where it
extracts one or more local level sets, removes any kinks in these by use
of reinitialization, and finally uses these local level sets to calculate the
curvature and normal vectors. The values corresponding to the interface which is
closest to the current grid point is used.

The method is motivated in that it is more general than the previous method by
Salac and Lu \cite{salac}, handling bodies which fold back onto themselves, and
it extends more easily to 3D than the previous methods by Macklin and Lowengrub
\cite{macklin,macklin06} and by Lervåg \cite{lervag-curv-conf,lervag-curv}, which
use curve-fitting schemes. The parameters of the method are given in
\cref{tab:parameters}. Results, both for static and dynamic simulations, are
given in the next sections.




\section{Geometric results}
\label{sec:geometric}
In order to test the LOLEX method, some static interface configurations were
used that replicate typical situations occurring in simulations of droplet
collisions.

\subsection{Circles and straight interfaces}
\begin{figure}[tbp]
  \centering
  \subfigure[LOLEX method]
  {\includegraphics[width=0.8\textwidth]{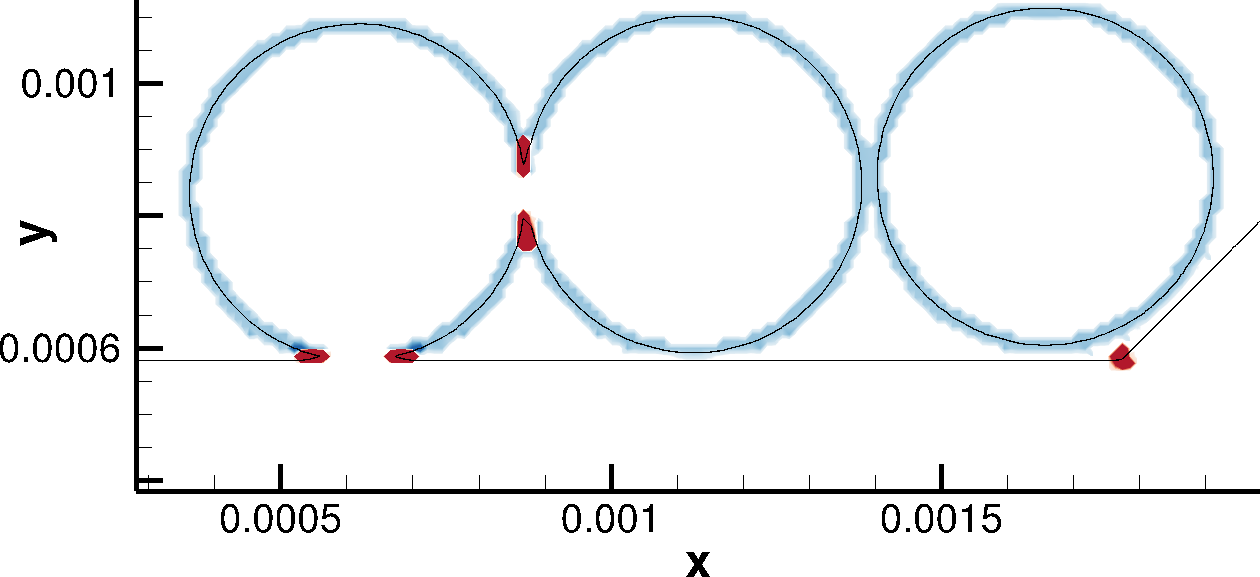}}
  \\
  \subfigure[Salac and Lu method]
  {\includegraphics[width=0.8\textwidth]{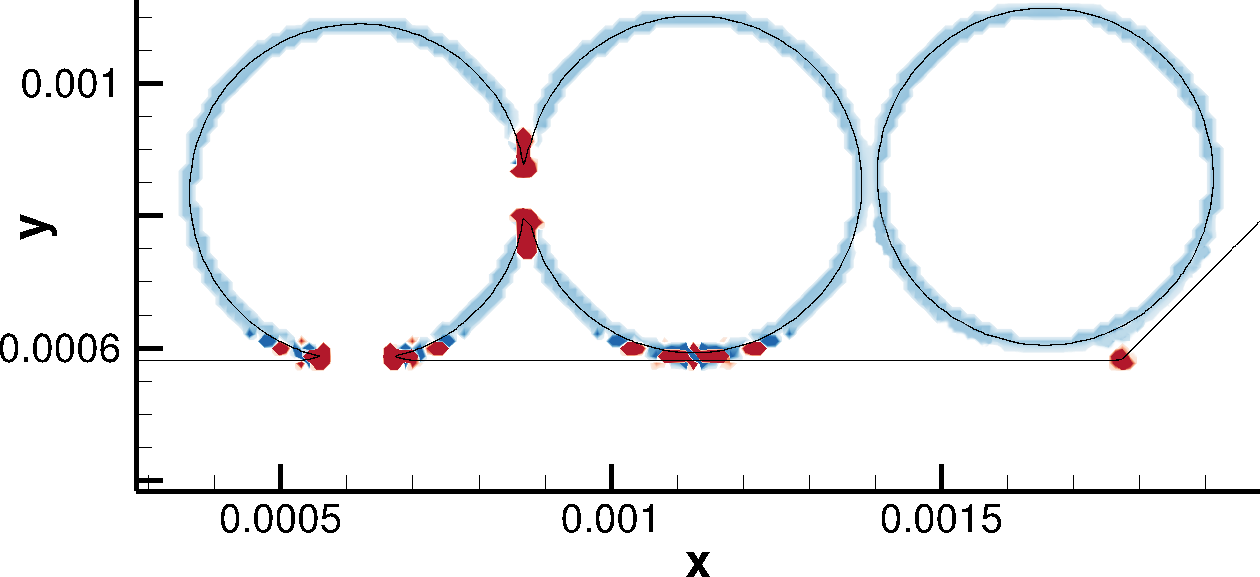}}
  \\
  \subfigure[Standard method]
  {\includegraphics[width=0.8\textwidth]{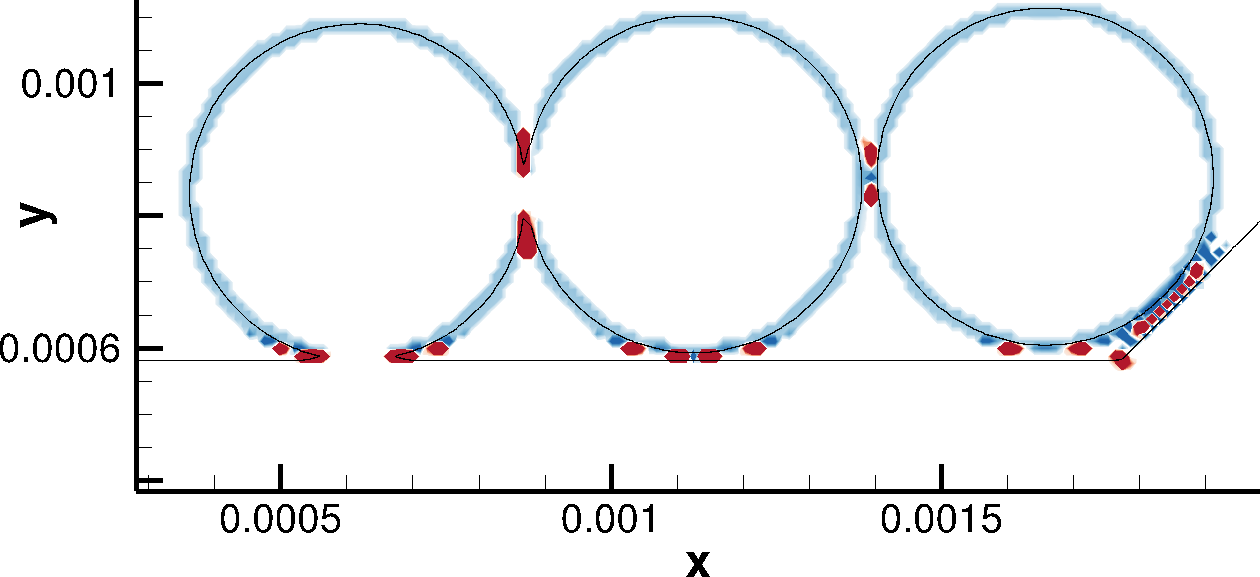}}
  \caption{Comparison of curvature calculation methods for circles and
    straight interfaces. The color indicates the curvature; white is zero,
    blue is negative and red is positive.}
  \label{fig:init-test}
\end{figure}

The first test case consists of three circles and a straight-lined interface,
where two of the circles and the straight-lined interface are joined together.
The results for this case are shown in \cref{fig:init-test} for the LOLEX
method, the SLM, and the standard method. In this figure, the interfaces are
shown as black lines, and the color indicates the curvature. The background
curvature of 0 is indicated in white, blue indicates a negative curvature and
red indicates a positive curvature.  The figure illustrates that the standard
method produces positive unphysical curvatures several places, both between the
circles and the straight interface and between circles. The Salac and Lu method
remediates the situation somewhat, but still has problems where the circle
folds back onto the straight interface, and at the bottom of the middle circle,
which is particularly close to the straight interface.  The LOLEX method
produces positive curvatures only where they are expected and needed.

\subsection{Droplet falling onto a pool}
\label{sec:drop-fall-test}
\begin{figure}[htbp]
  \centering
  \includegraphics[width=\textwidth]{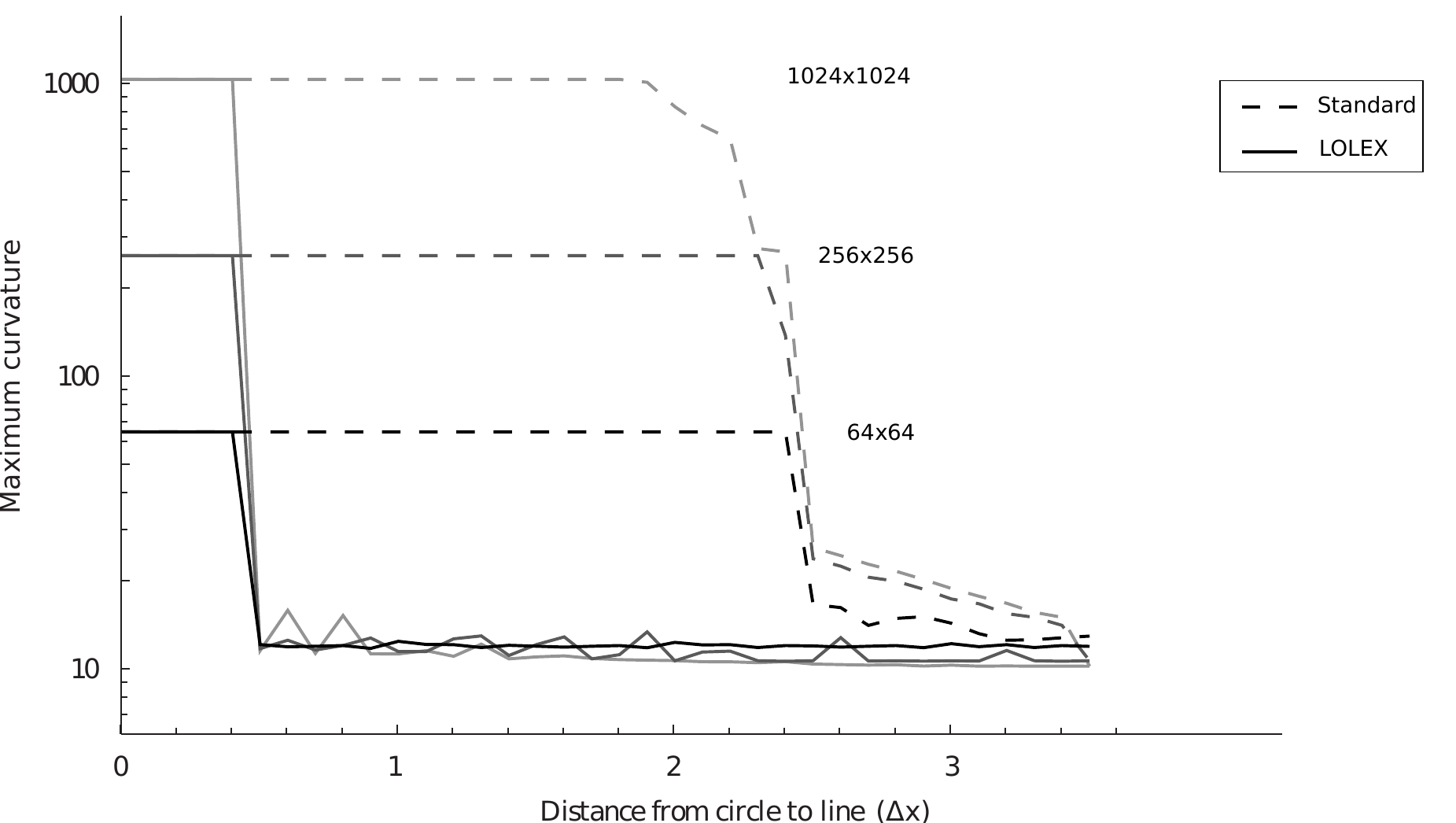}
  \caption[Supremum norm of curvatures]{Supremum norm of the curvature for
    decreasing interface separation. Dashed lines: results using the
    standard method. Solid lines: results using the LOLEX method. The lines
    are shaded lighter with increasing grid resolution. The analytical
    curvature of the circle is 10.}
\label{fig:curv-inf-norms}
\end{figure}

In order to compare the behavior of the LOLEX and the standard method for
different interface separations, a test case was considered which mimics
a droplet falling onto a pool. In this case, a 0.2 m diameter circle and
a horizontal line were initialized in a 1m \x 1m domain.
The separation between
the circle and the line was varied from $3.6\Delta x$ down to $0\Delta x$ in increments of $0.1\Delta x$.
For each separation, the curvature was calculated at all points
within the narrow band close to the interfaces, and the supremum-norm
$\|\kappa\|_{\infty}$ of the curvature values was calculated. This was done using the
standard and the LOLEX method, for grid resolutions of 64\x64, 256\x256 and
1024\x1024. The analytical curvature is 10 for the circle and 0 for the line, so
the supremum norm should be close to 10. The results are shown in \cref{fig:curv-inf-norms}.

As is seen in this figure, the standard method returns the value used in
regularizing the curvature, $\|\kappa\|_{\infty} = \frac{1}{\Delta x}$, when the
interface separation becomes smaller than about 2.4 $\Delta x$. Increasing the
grid resolution does not improve the situation. Note that the $y$ axis in this
plot is logarithmic. Meanwhile, the LOLEX method gives decent values close to
the analytical value of 10 all the way up to when the
interfaces merge, which happens at a separation of 0.2 $\Delta x$. It is seen
that the small deviations for the LOLEX method are reduced when the grid
resolution is increased.

A final thing which may be illustrated using this figure is the effect of the
parameter $\eta$. This parameter indicates how much the level-set function
$\phi$ has to deviate from being a signed distance function before we switch
from using the standard method to the LOLEX method. In the circle-line case in
\cref{fig:curv-inf-norms} the value of $\eta = 0.005$ (used throughout this
paper) triggers the switch for the first point when the distance is 
$4.2 \Delta x$. Using $\eta = 0.01$ as in \cite{macklin06}, the switch is made
at $3.9 \Delta x$. Both these distances are in the region where the standard
method gives good answers, so the LOLEX method is not very sensitive to the
precise value of $\eta$ as long as it is in this range.

In addition to the curvature, accurate normal vectors close to the interface are
desirable in level-set simulations. The importance in reinitialization has been
suggested above, coming from the fact that normal vectors are used in finding
the upwind direction. Normal vectors are equally important in calculating the
extension velocity, where an error would lead to the interface not moving
according to the flow.

\subsection{Concentric circles}
\begin{figure}[htbp]
        \begin{center}
                \subfigure[Central difference]
          {\includegraphics[width=0.44\textwidth]
      {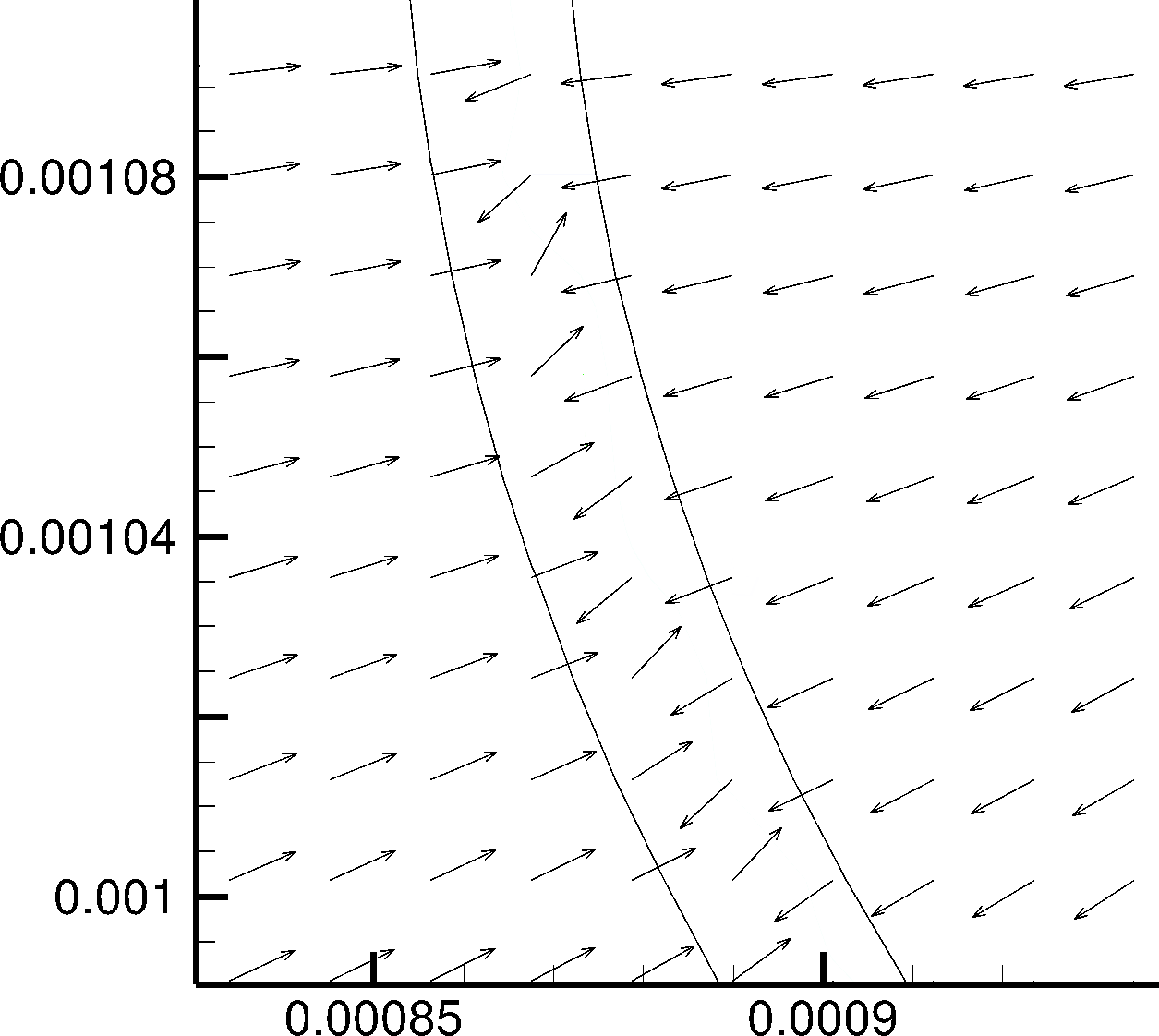}}
    $\quad$
                \subfigure[Directional difference]
                {\includegraphics[width=0.44\textwidth]
      {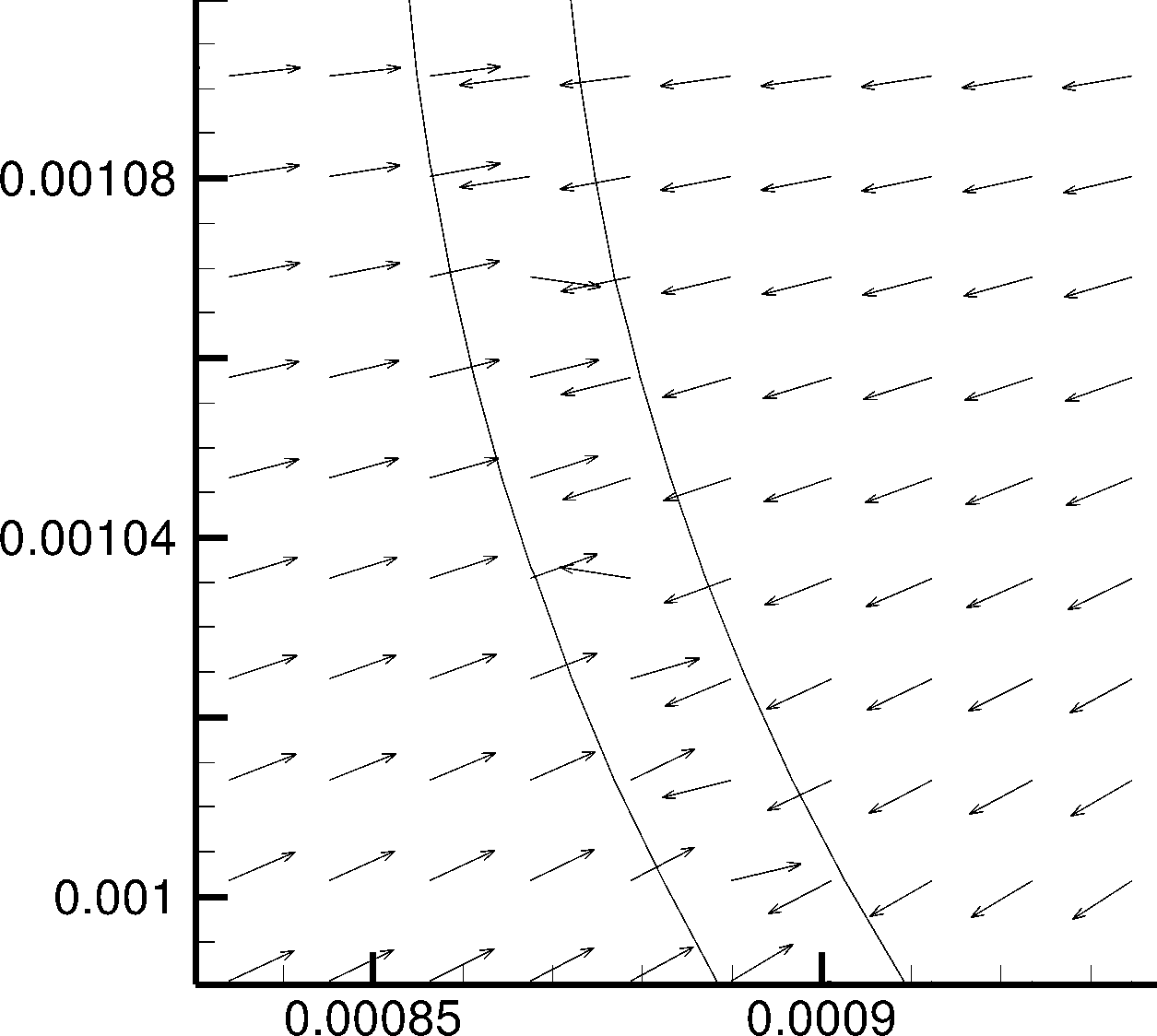}}
    \\
                \subfigure[Lervåg]
                {\includegraphics[width=0.44\textwidth]
      {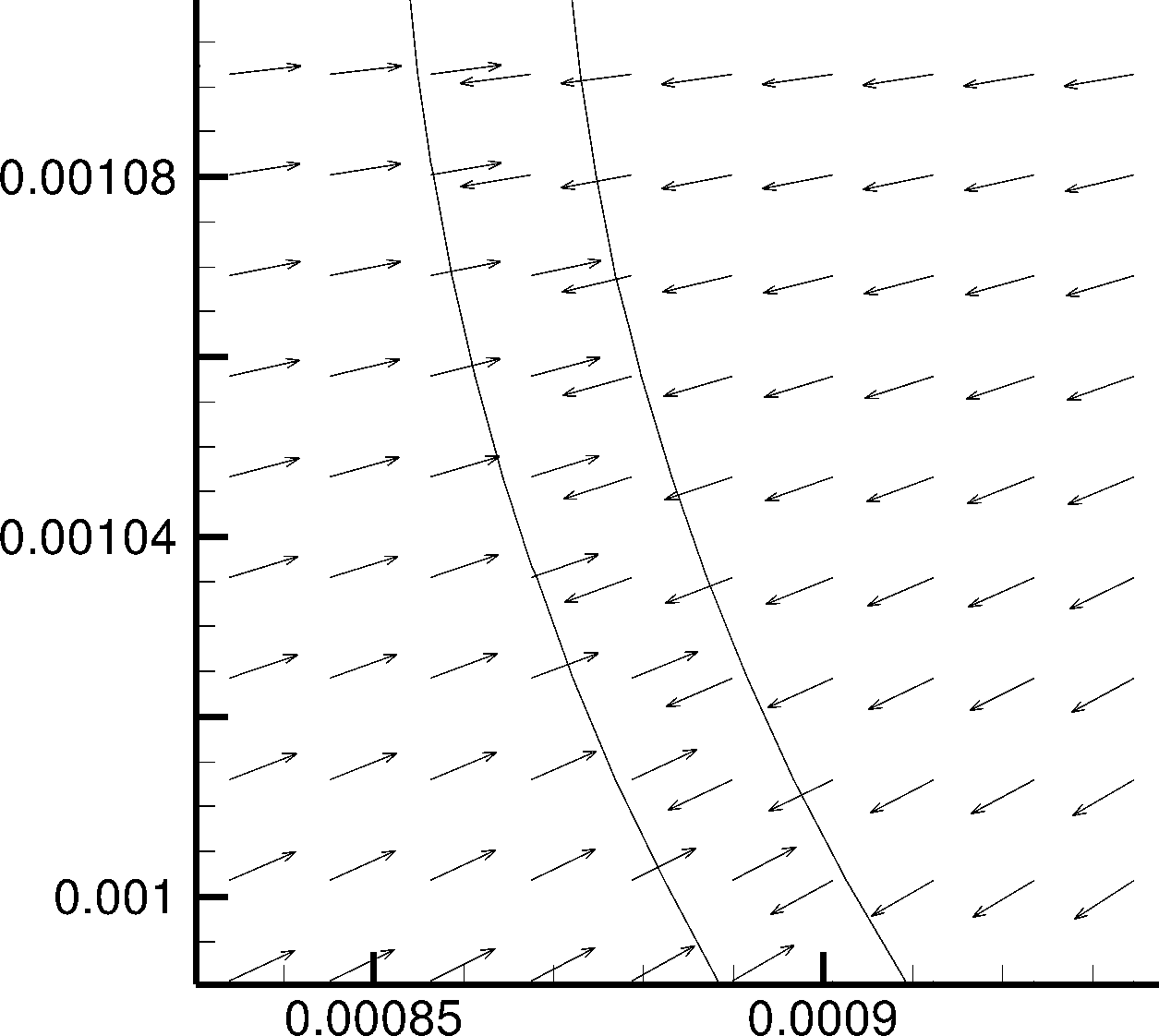}}
    $\quad$
                \subfigure[LOLEX]
                {\includegraphics[width=0.44\textwidth]
      {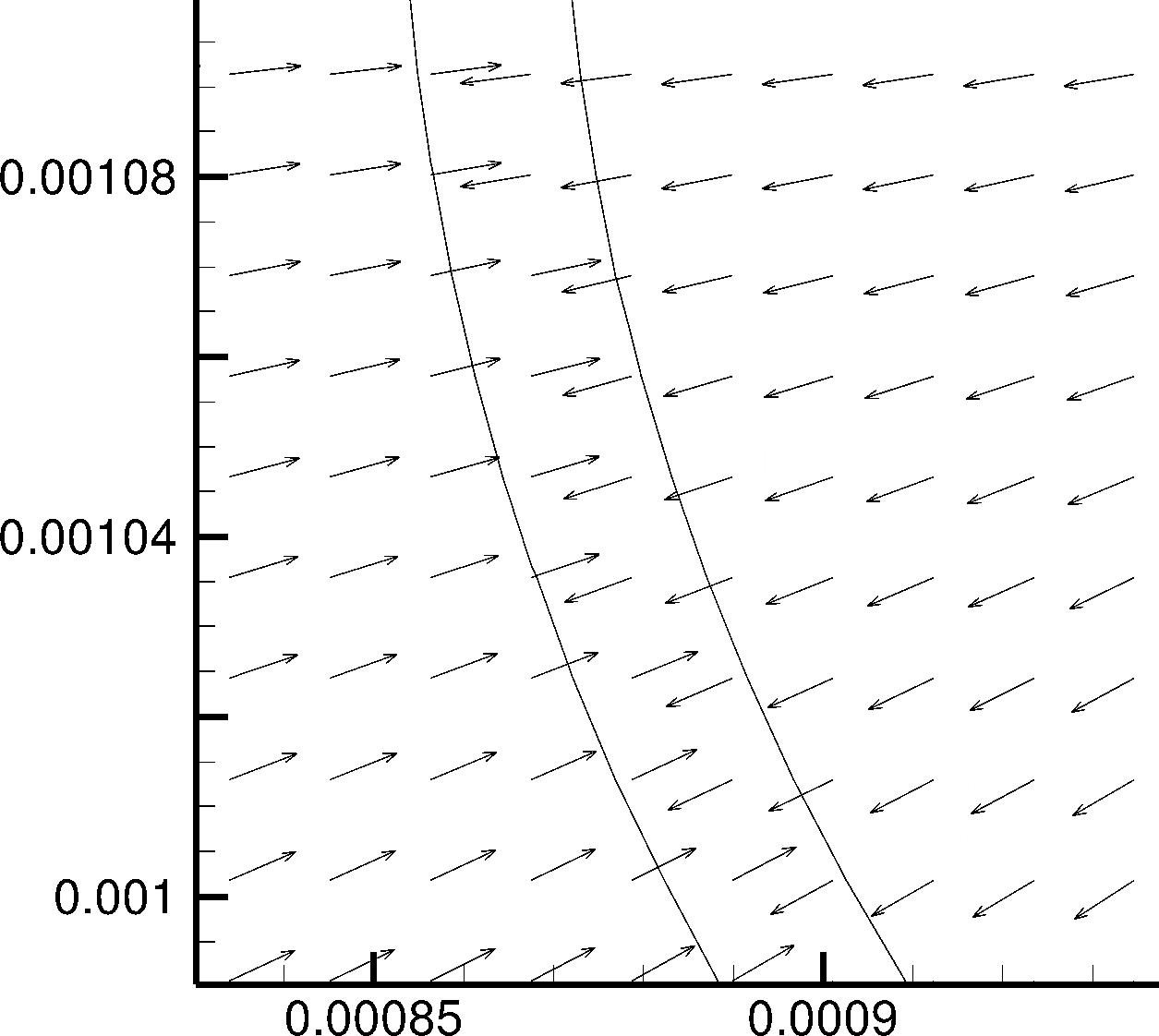}}
  \end{center}
        \caption[Comparison of normal vectors]{Comparison of normal vector
  calculations using different methods.}
\label{fig:normals}
\end{figure}
In order to compare the proposed method to the standard method, a geometric test
case was considered which replicates the demands of simulating merging
interfaces. The calculated normal vectors are compared both to the
standard central-differences discretization, to a directional-differences
discretization as described in \cite{macklin}, and to the curve-fitting
method of Lervåg \cite{lervag-curv}.

In this test case, two concentric circles were initialized, as if we had a thin
ring of fluid 1 inside fluid 2. The width of this ring was $1.6 \Delta x$. This
test case is interesting, since it reveals grid effects or anisotropies. It
also replicates the situation of a thin film that forms between a droplet and
a pool for cases where the droplet deforms the pool surface before merging. This
has been observed experimentally, see \eg \cite{longmire-fall}. The results for
all four methods are shown in \cref{fig:normals}.

In this figure it is seen that the directional difference
method is not much better than the central difference method. This is partly what
prompted the use of curve fitting methods; Macklin and Lowengrub initially used
directional differences and additional grid refinement in \cite{macklin}, but
switched to curve-fitting methods in \cite{macklin06}.
As is seen in \cref{fig:normals} (c), curve fitting
methods (the method by Lervåg is used here) give the correct result. In (d), we
see that the LOLEX method also gives the correct result. It is impossible to
distinguish the results in (c) and (d) without overlaying the figures and
zooming in a lot. We calculated the maximum RMS deviation between the
LOLEX method and the other three methods compared in \cref{fig:normals}, \eg $\|
\sqrt{(n_x(d) - n_x(a))^2 + (n_y(d) - n_y(a))^2}\|_{\infty}$. This was 0.0086
for the Lervåg method, 0.92 for the Central difference and 1.78 for the
Directional difference; a 90$^{\circ}$ difference would give $\sqrt{2}$. We note
that the maximum error is largest for the Directional difference, while the
average error is largest for the Central difference.
The difference between the LOLEX and Lervåg methods is too small to have any 
impact on the simulation results. 

As pointed out several times already, the main advantage
of the present LOLEX method over methods employing curve fitting is that it
scales easily to 3D. This is because the present method retains the implicitness
of the level-set method. A 3D extension of the Macklin and Lowengrub
method, on the other hand, would fit a local surface to the point of interest,
as they indicate in \cite{macklin06}.
Curvature estimation in 3D based on local surface fitting has long been a
topic of research in computer vision, see \cite{surface-fitting} for a review
of various methods including the use biquadratic surfaces and of splines. The
conclusion of \cite{surface-fitting} is that these methods are very sensitive to
numerical noise (in their context, sensor noise). In the current case, noise is
to be expected, as can be seen in \cref{fig:reinit-lower} (b). Due to this fact,
methods in computer vision that avoid local surface fitting and calculate only
the sign of the curvature have been introduced, since this quantity can be
calculated more reliably\cite{no-surface-fitting}. This is not a viable
alternative in two-phase flow simulations as considered here.

\subsection{3D bubble above a plane}
\begin{figure}[tbp]
        \noindent\makebox[\textwidth]{%
    \subfigure[Standard discretization]{
      \includegraphics[width=0.6\linewidth]
        {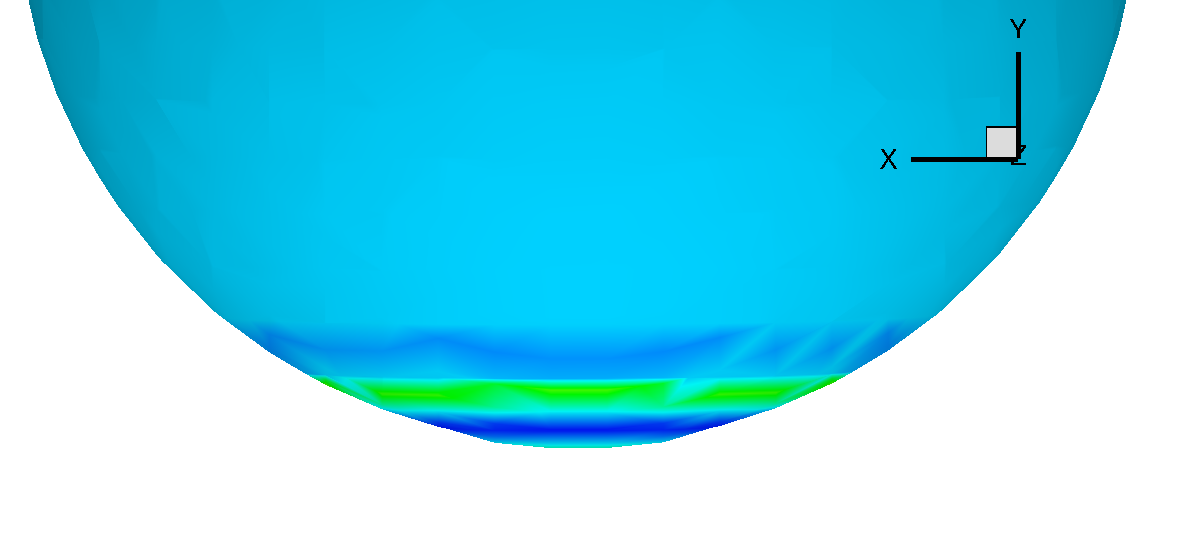}
    }
    \subfigure[LOLEX method]{
      \includegraphics[width=0.6\linewidth]
        {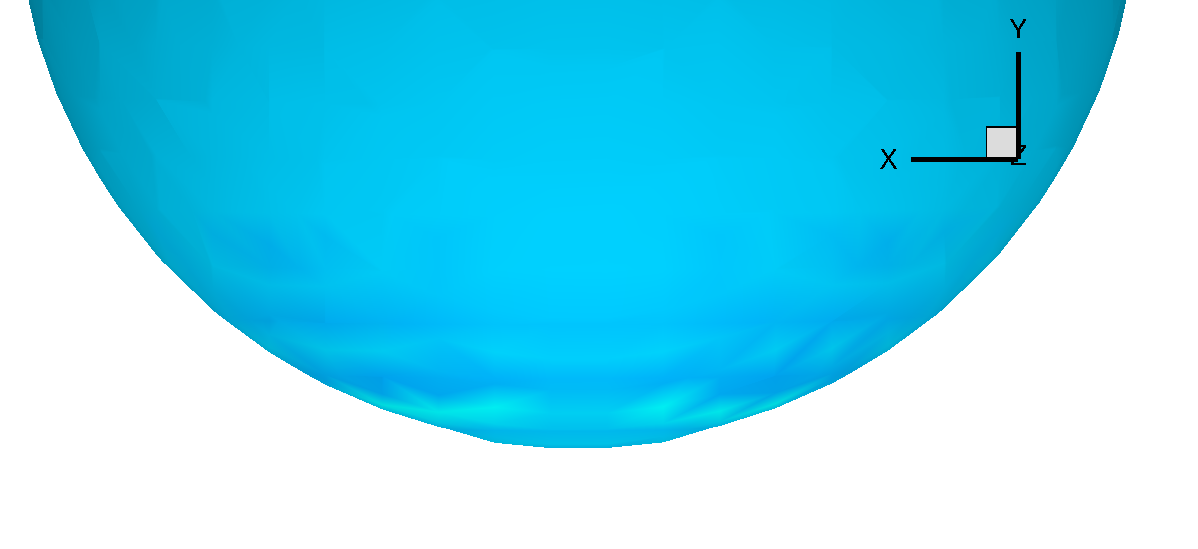}
    }
  }%
  \caption[3D results of the LOLEX method compared to the standard method]%
  {3D bubble above a plane (not shown). Comparison of the standard
    curvature discretization (a) and the LOLEX method (b). The surfaces are
    colored according to the curvature, and the standard method is seen to
    give large errors close to the kink in the level-set function (which is
    below the sphere), seen as green and dark blue bands.}
  \label{fig:3d}
\end{figure}
A curvature calculation using
the LOLEX method on a 3D case is shown in \cref{fig:3d}.  In this case, a bubble
is placed above a plane, with distance 1.2 $\Delta x$ at the closest. The grid is
50\x50\x50, and the bubble radius is 12.5 $\Delta x$.  The surfaces are colored
according to the curvature (interpolated to the surface). In \cref{fig:3d} (a),
the standard method is used. In 3D, this is the 27-point stencil given by Kang et
al. \cite{kang}. In \cref{fig:3d} (b), the LOLEX method is used to extract the
local level sets, and the curvature is then calculated using the same 27-point
stencil on these local level sets. It is seen that the LOLEX method performs
much better than the standard discretization in areas where the bubble and plane
are in close proximity. Note that the plane is not shown here, only the bubble.
The kink in the global $\phi$ is below the bubble.

Comparing to the analytical curvature, which in this case is $-10$ for the
spherical bubble, it is seen that the standard discretization performs well away
from kinks, where the variation in curvature is at most $\pm 0.2\%$. Close to
the kink, the standard discretization has errors of $\pm 80\%$, seen as green
and dark blue bands in \cref{fig:3d} (a). The LOLEX method has the same
variation as the standard method away from kinks, while the variation is $\pm
2\%$ close to the kink, seen as light blue spots in \cref{fig:3d} (b).  Thus it
is seen that the LOLEX method gives an error which is an order of magnitude
lower than the standard method close to kinks in the level-set function. There
is still a small error of the same size as reported above in 2D, which
is again probably caused by reinitialization. A deviation of this
magnitude is unlikely to have a large impact on simulations, in contrast to the
errors from the standard discretization.

To the knowledge of the authors, improved calculation of geometric quantities
for a pure level-set formulation in three spatial dimensions that handles
general topologies have not been reported before in the literature. Salac and Lu
report results of 3D simulations in \cite{salac}, but it is not known how (or
if) they handle problems like that illustrated in \cref{fig:salac-problem}, \ie
a body folding back onto itself.  They also do not discuss the problem of
needing good normal vectors at the interface in order to solve the
reinitialization equation.

Given the current state of developments toward petascale supercomputers,
and particularly the rapid evolution in GPU-accelerated solvers, dynamic 3D
level-set simulations of colliding bodies are going to become more and more
common. As this happens, a method such as the present one will be necessary
in order to get trustworthy results for situations where accurate curvature is
important.


\section{Dynamic results}
\label{sec:simulations}
As discussed previously, the case of a single droplet of liquid falling onto
a pool of the same liquid, either through gas or another liquid, has been
widely studied.  Thus it is a good benchmark on which to test the proposed
method, since detailed experimental results are available.

When considering this case, the main dichotomy is between a droplet falling
through gas and a droplet falling through liquid.  We will consider both cases
here, since both are interesting from an industry standpoint.
These two cases present different challenges to numerical simulations.  The
liquid-in-gas case has a high density difference between the two fluids, which
is known to be a difficult case. Sussman \etal have studied this problem, and
have produced good results using a hybrid LSM-VOF method \cite{hybridlsmvof}.
The liquid-in-liquid case, on the other hand, can be time-consuming to simulate
due to the viscous term in the CFL-criterion used here \cite{kang}, but is not
challenging with respect to density differences.

\subsection{Decane droplet in water merging with decane pool}
\label{sec:decane-exp}
The simulation discussed here consider two immiscible liquids, where
a droplet of the heaviest liquid is placed in the lightest liquid above a pool
of the heaviest liquid. In the experimental work by Chen \etal
\cite{water-decane}, the droplet is made to rest on the pool, and then merging
happens after some time. The heavy liquid is water, and the
light liquid is a mix of 20 \% polybutene in decane. The droplet diameter is 1.1
mm. As the droplet and interface are brought into proximity, a thin film is
formed between them. This thin film drains, and after some time the film
ruptures and the droplet merges with the interface. In the Chen \etal
experiments, the merging happens at the central point, but off-center merging
has also been reported for larger droplets \cite{longmire}.

\begin{figure}[tbp]
  \centering
  \subfigure[Experimental result]
    {\includegraphics[width=\linewidth]{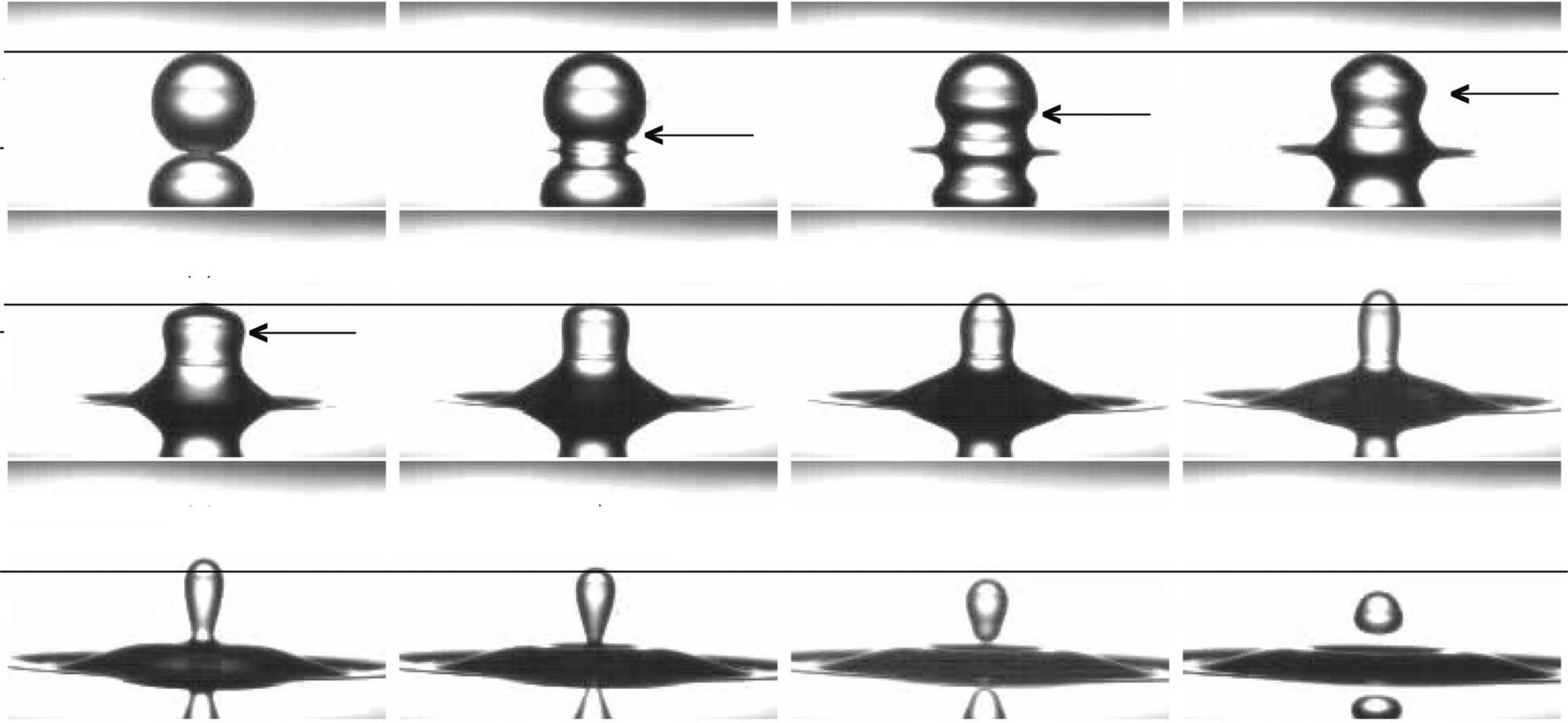}}
  \\
  \subfigure[Simulation result]
    {\includegraphics[width=\textwidth]{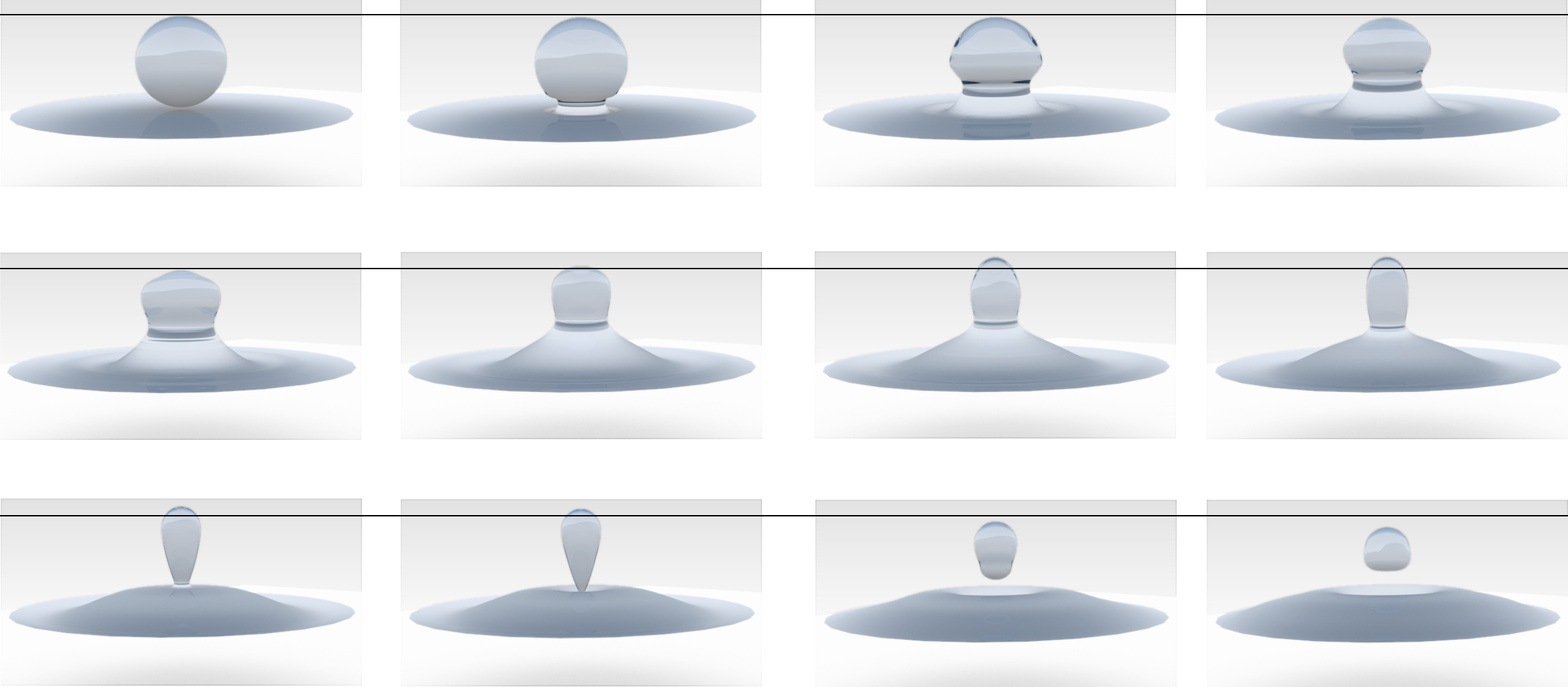}}
  \caption{A 1.1 mm diameter water droplet merging with a water pool. The
    ambient fluid is 20\% polybutene in decane. Snapshots are taken 542
    $\mu$s apart, the arrow indicates the capillary wave, and the
    horizontal lines indicate the top of the bubble in the first
    frame. Figure (a) is the experimental result, reprinted with permission
    from \cite{water-decane}, copyright 2006 American Institute of
    Physics. Figure (b) is the simulation result. }
  \label{fig:chen-exp}
\end{figure}

A simulation was performed with the same fluid properties and droplet dimension
as reported by Chen \etal The computational domain was 6\x6 mm, the grid
was 400\x400, and the CFL-number was 0.8. The results are shown in \cref{fig:chen-exp}.
The agreement between the simulation and experimental results is very good.

In this simulation, the point of merging is decided mainly by grid effects when
the droplet deforms the interface forming a thin film.  With the present method,
we must simply hope that precisely what happens at the time of merging does not
significantly affect the following behaviour. Comparing \cref{fig:chen-exp} (a)
and (b) indicates that in this case the precise mechanisms of merging are not
very important, as the numerical and experimental results agree very nicely. To
accurately capture the thin film behaviour, the grid resolution would have to be
extremely fine. Hodgson and Lee \cite{surfactants} report that the width of the
thin film between a droplet and a pool for the water-toluene system they study
is four orders of magnitude smaller than the droplet diameter, \ie around 100
nanometers. It is possible that an adaptive grid could be able to resolve such
a thin film, but since there is no analog to the Knudsen number for liquids, it
is not immediately clear whether the continuum description of the Navier-Stokes
equations still holds at this length scale.

\begin{figure}[tbp]
  \centering
    \includegraphics[width=0.6\linewidth]{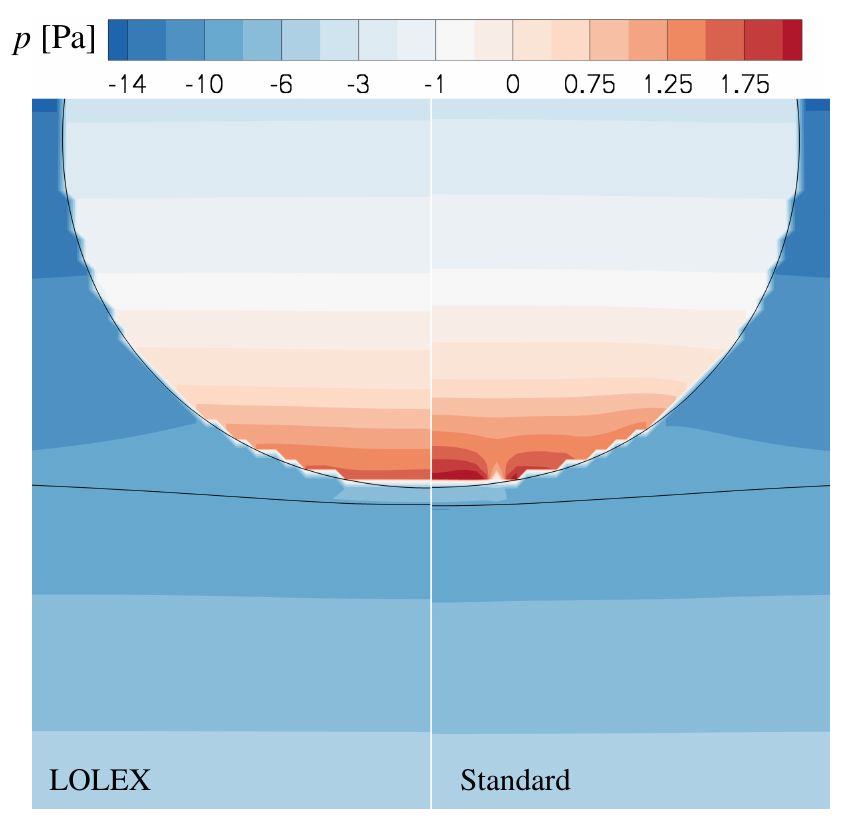}
  \caption{Water droplet in a mixture of polybutene and decane, about to
    merge with a water pool. Comparison of the pressure field using the
    LOLEX method and the standard method at $t=0.007$ s.}
  \label{fig:wd-pressure}
\end{figure}

Comparing the LOLEX method and the standard method on this case, the standard
method gives a more oscillatory pressure field around the contact point, as
seen in \cref{fig:wd-pressure}.  This increased pressure inside the thin film
delays the rupture of the film, seen as a slightly increased width of the film in
\cref{fig:wd-pressure} (b).

\subsection{Water droplet falling through air onto a water pool}
Considering the case of a liquid in gas, a simulation was performed of a 0.18
mm diameter water droplet falling through air at 0.29 m/s and impacting a deep
pool of water.  Experimental results for this case due to Zhao, Brunsvold and
Munkejord are found in \cite{zhao-multiphase}.  These results indicate that
a partial coalescence occurs, but the high-speed camera used was not fast
enough to capture all the details of the partial coalescence process.

\begin{figure}[tbp]
  \centering
  \subfigure[Experimental result]
    {\includegraphics[width=\linewidth]{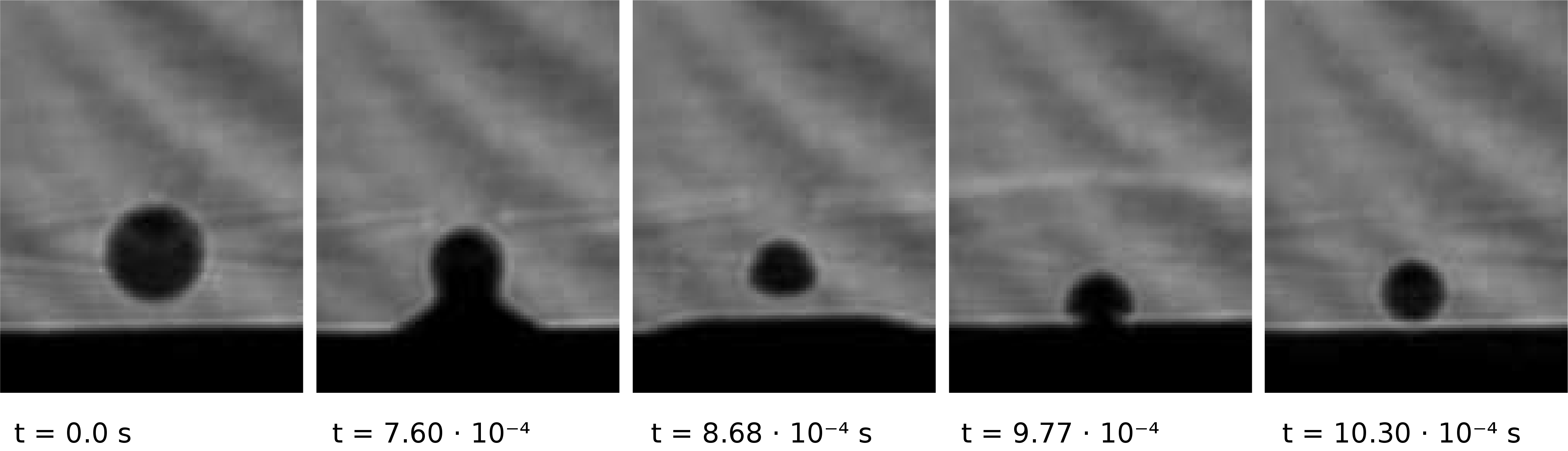}}
  \\
  \subfigure[Simulation result]
    {\includegraphics[width=\linewidth]{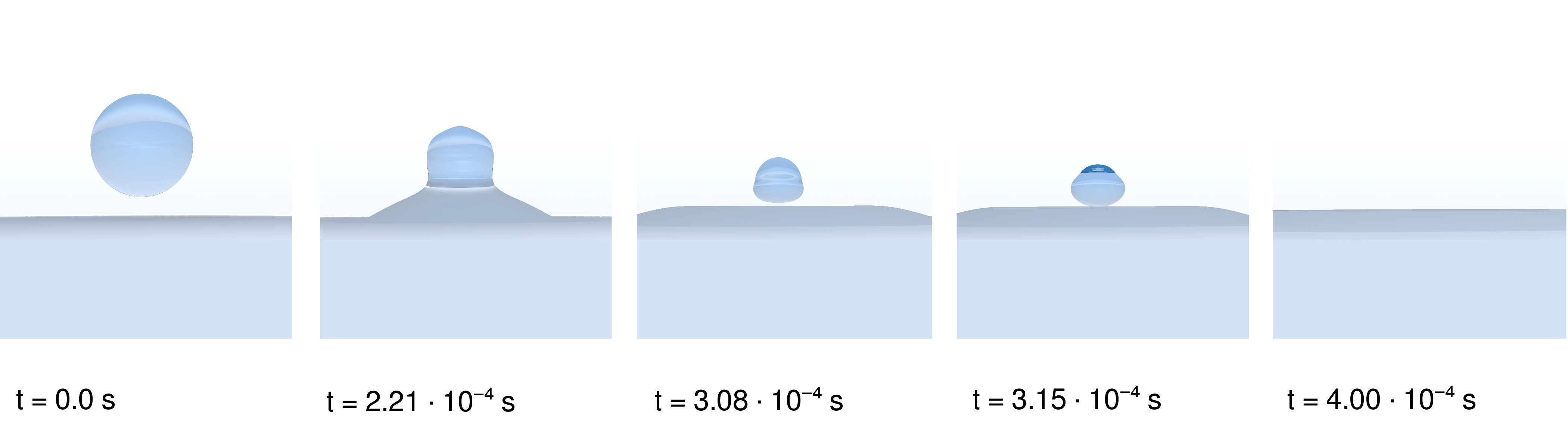}}
  \caption{Experimental results (top) and simulation results (bottom) for
    a 0.18 mm water droplet falling through air and impacting a deep pool of
    water at 0.29 m/s.  Figure (a) is reprinted from \cite{zhao-multiphase},
    Copyright (2011), with permission from Elsevier.}
  \label{fig:water-air-comp}
\end{figure}

The simulation was performed using axisymmetry.  The computational domain was
0.7~mm~\x~0.7 mm, resolved using a 401\x401 Cartesian grid. The CFL number was
0.25. The LOLEX method was used for curvature and normal vector calculation.
A comparison of the experimental and simulation results are shown in
\cref{fig:water-air-comp}.

The time intervals between frames for the experimental and simulation results do
not match in this figure. The intervals between the second and third frames are
the ones that match best, suggesting that the behaviour of the thin air film
that forms between the droplet and the pool before coalescence is the major
source of this discrepancy. The grid used in the simulation is unable to
resolve the thin film. It is not clear that an increased grid resolution would
amend this, as the continuum approximation may not be valid for the thin air
film. The width of this film is not known from experiments.

As an order-of-magnitude estimate, we can use the results by Hodgson and Lee
\cite{surfactants}.  They report that the width of the thin film between
a droplet and a pool before merging, for the water-toluene system they study, is
around $L=100$ nanometers.  Since the mean free path in air at room temperature
and atmospheric pressure is around $\lambda = 66$ nanometers
\cite{meanfreepath}, the Knudsen number is $Kn = \frac{\lambda}{L} \approx 0.7
\centernot\ll 1$, which would imply that the continuum description is no longer
valid.

Nevertheless, the simulation is able to correctly predict the partial
coalescence, and the simulation agrees well with experiments on the size of the
daughter droplet produced. In the experiments, this daughter droplet
subsequently bounces on the pool of water.  The simulation is unable to predict
this, again due to the thin air film formed, and shows the daughter droplet
merging with the water pool instead.

\begin{figure}[htbp]
  \centering
  \subfigure[Reinitialization every 7 time steps]{
        \includegraphics[width=0.47\textwidth]{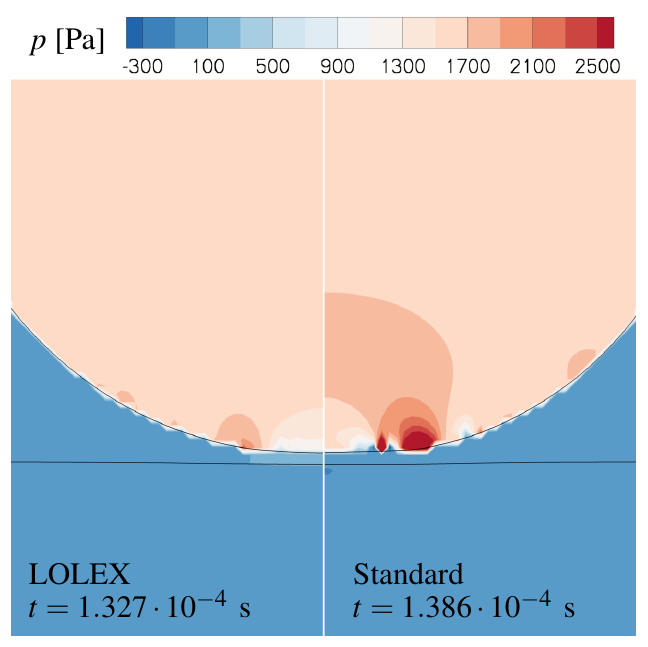}}
  \subfigure[Reinitialization every time step]{
        \includegraphics[width=0.47\textwidth]{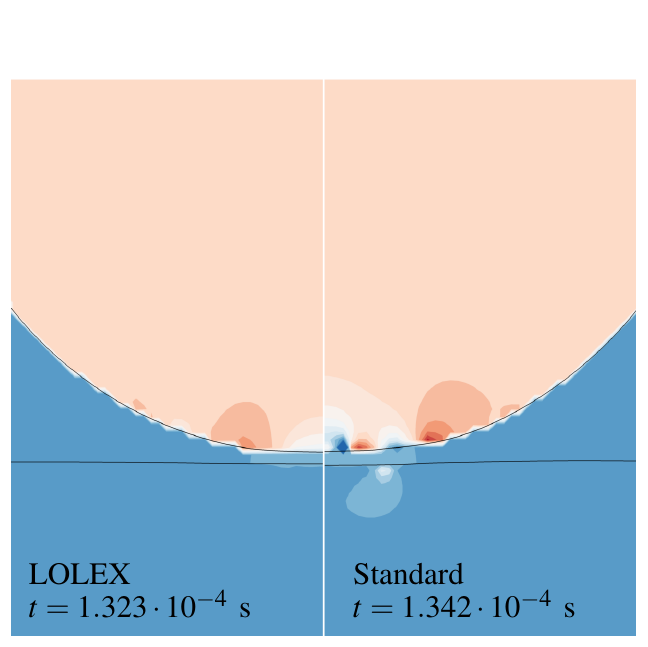}}
  \caption{Water droplet falling onto a pool, just before the interfaces
    merge. Comparison between the LOLEX method and the standard method.
    The pressure field is shown as colored contours.}
  \label{fig:lolex-pres_coll}
\end{figure}

\begin{figure}[htbp]
  \centering
  \subfigure[Reinitialization every 7 time steps]{
    \includegraphics[width=0.47\textwidth]{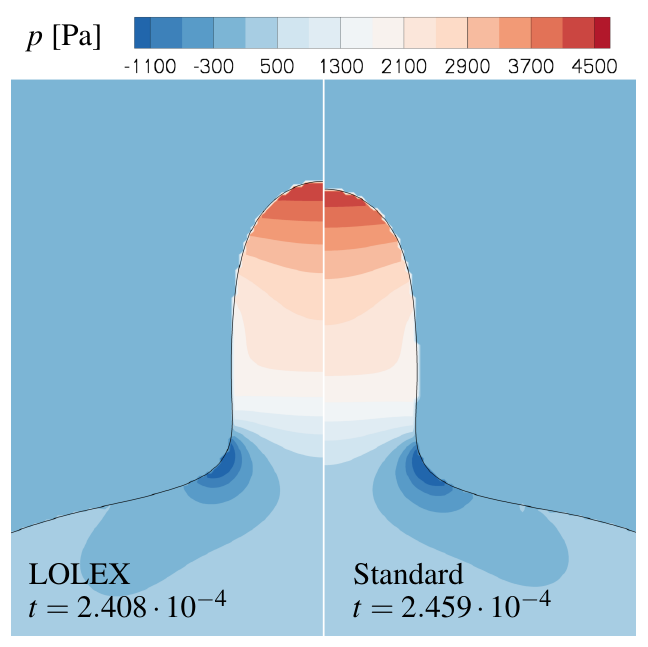}}
  \subfigure[Reinitialization every time step]{
    \includegraphics[width=0.47\textwidth]{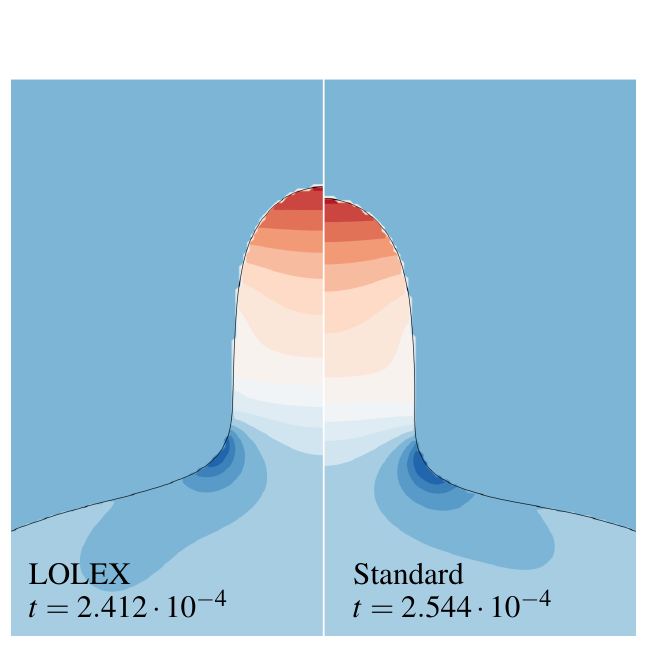}}
  \caption{Water droplet falling onto a pool, when the neck reaches its
    highest position. Comparison between the LOLEX method and the standard
    method. The pressure field is shown as colored contours.}
  \label{fig:lolex-pres_neck}
\end{figure}

A comparison between the LOLEX method and the standard method is shown in
\cref{fig:lolex-pres_coll,fig:lolex-pres_neck}.  These figures show a section
through the droplets just before collision and just when the neck is at its
tallest, respectively.  The pressure field is plotted as colored contours.  The
LOLEX method is plotted on the left side and the standard method is plotted on
the right side.  It is seen from these figures that the curvature errors
produced by the standard method give rise to significant oscillations in the
pressure; note in particular the interleaved red and blue patches where the
pressure changes sign. As the reinitialization is performed more frequently, 
the oscillations persist, and are even found inside the pool below the droplet.

An important effect of this erroneous pressure is a loss of
kinetic energy, which can be seen in \cref{fig:lolex-pres_neck}, where the
neck is clearly shorter with the standard method. It is also seen that
more frequent reinitialization leads to a higher loss of kinetic energy. As some
authors have noted \cite{blanchette}, the height of the neck and the dynamics of
the capillary waves are important factors for the partial coalescence mechanism.

The LOLEX method is not significantly affected by the amount of
reinitialization, and gives a more sensible pressure field in both
cases. It should be noted that the pressure difference across the droplet
interface in \cref{fig:lolex-pres_coll} is about 2500 Pa, which is very large,
caused by the very small droplet diameter. 

\begin{figure}[htbp]
  \centering
  \includegraphics[width=\linewidth]{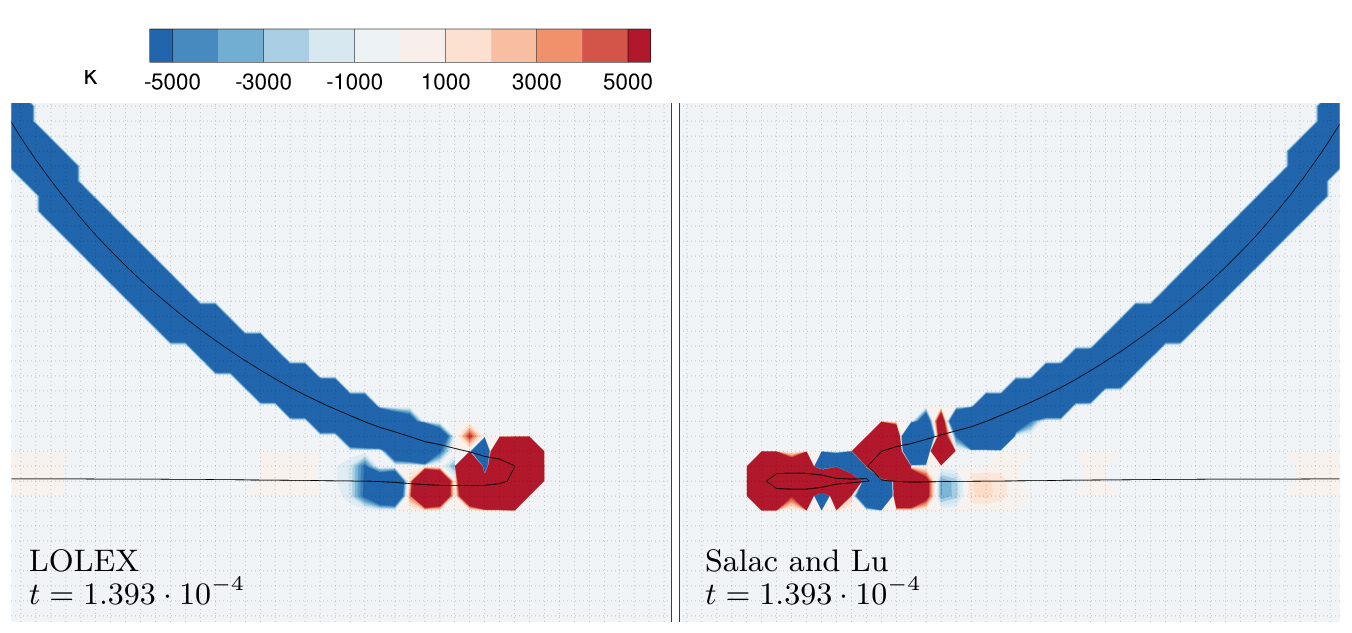}
  \caption{Water droplet falling onto a pool, zoom in on the interface just
    after merging. Comparison between the LOLEX method and the Salac and Lu
    method. It is seen that the latter entrains a small air bubble due to the
    oscillatory curvature field following the merging. The curvature field is
    shown as colored contours.}
  \label{fig:lolex-slm-comp}
\end{figure}

This case also allows an illustration of the benefit of using the LOLEX method
over the Salac and Lu method. In \cref{fig:lolex-slm-comp} we compare the curvature
field for these two just after droplet-film merging, where it is
seen that curvature errors in the Salac and Lu method have led to the
entrainment of a small bubble. Since the bubble is under-resolved on this grid,
it subsequently disappears due to reinitialization. The LOLEX method does not
entrain any bubbles.

Finally, we consider the performance impact of the LOLEX method on this case.
The same simulations using the LOLEX and standard method were rerun using
a 201\x201 grid for timing purposes. The simulation using the standard method
took 43525 s of CPU time, while the simulation using the LOLEX method took 46753
s. This means the LOLEX method is 7\% slower than the standard method for this
case, which is a fair trade-off for the benefits of both reduced pressure
oscillations and lower sensitivity to reinitialization frequency.

\section{Concluding remarks}
\label{sec:conclude}
In the present work we have proposed a new method for calculating the
curvature and normal vectors of an interface represented by a level-set
function, and which gives accurate results before, during and after
topological changes in the interface. The method is compared to the
standard method for geometric test cases, where the analytical curvature is
known, and it is seen that in areas where the standard method gives errors
of around 100 \%, the proposed method gives errors of around 1--2 \%. The
method is easily extended to 3D, as is demonstrated, where the same
reduction in error is seen. The method is then employed in simulations of
two-phase flow where a droplet merges with a pool.  Here it is seen that
the standard method gives rise to unphysical pressure oscillations before
merging, which affect the subsequent capillary waves, while the proposed
method fares much better. The results of the simulations using the proposed
method are compared to experimental results both for a liquid-in-liquid
case, where the agreement is very good, and for a more demanding
liquid-in-gas case where the agreement is qualitative, reproducing the
partial coalescence behaviour.


\section*{Acknowledgements}
This work was financed through the Enabling Low-Emission LNG Systems project at
SINTEF Energy Research, and the authors acknowledge the contributions of
GDF SUEZ, Statoil and the Petromaks programme of the Research Council of
Norway (193062/S60). We also thank one of the anonymous reviewers for detailed
and extensive comments which helped improve the quality of this work.

\bibliographystyle{model1-num-names}
\bibliography{1-article}

\end{document}